\documentclass[%
reprint,
superscriptaddress,
amsmath,amssymb,
aps,
prb,
]{revtex4-2}

\usepackage{graphicx}
\usepackage{dcolumn}
\usepackage{bm}
\usepackage{braket}
\usepackage{appendix}
\usepackage{subfigure}
\usepackage{hyperref}
\usepackage{color}
\usepackage{xcolor}
\usepackage{amsmath}
\usepackage{nccmath}
\usepackage[english]{babel}
\usepackage{svg}

\begin{document}

	\title{Quantum sensing and imaging with spin defects in hexagonal boron nitride}
	\author{Sumukh Vaidya}
	\affiliation{
		Department of Physics and Astronomy, Purdue University, West Lafayette, Indiana 47907, USA
	}
	\author{Xingyu Gao}
	\affiliation{
		Department of Physics and Astronomy, Purdue University, West Lafayette, Indiana 47907, USA
	}
	
	\author{Saakshi Dikshit}%
	\affiliation{%
		Elmore Family School of Electrical and Computer Engineering, Purdue University, West Lafayette, Indiana 47907, USA
	}%
	
	\author{Igor Aharonovich}%
	\affiliation{
		School of Mathematical and Physical Sciences, University of Technology Sydney, Ultimo, New South Wales 2007, Australia
	}
	\affiliation{%
		ARC Centre of Excellence for Transformative Meta-Optical Systems, University of Technology Sydney, Ultimo, New South Wales 2007, Australia
	}
	
	\author{Tongcang Li}%
	\email{tcli@purdue.edu}
	\affiliation{
		Department of Physics and Astronomy, Purdue University, West Lafayette, Indiana 47907, USA
	}
	\affiliation{%
		Elmore Family School of Electrical and Computer Engineering, Purdue University, West Lafayette, Indiana 47907, USA
	}
	\affiliation{
		Purdue Quantum Science and Engineering Institute, Purdue University, West Lafayette, Indiana 47907, USA
	}
	\affiliation{
		Birck Nanotechnology Center, Purdue University, West Lafayette, Indiana 47907, USA
	}
	\date{\today}

\begin{abstract}
Color centers in hexagonal boron nitride (hBN) have recently emerged as promising candidates for a new wave of quantum applications. Thanks to hBN's high stability and 2-dimensional (2D) layered structure, color centers in hBN can serve as robust quantum emitters that can be readily integrated into nanophotonic and plasmonic structures on a chip. More importantly, the recently discovered optically addressable spin defects in hBN provide a quantum interface between photons and electron spins for quantum sensing  applications. The most well-studied hBN spin defects, the negatively charged boron vacancy ($V_B^-$) spin defects, have been used for quantum sensing of static magnetic fields, magnetic noise, temperature, strain, nuclear spins, {\color{black} paramagnetic spins in liquids}, RF signals, and beyond. In particular, hBN nanosheets with spin defects can form van der Waals (vdW) heterostructures with 2D magnetic or other  materials for in situ quantum sensing and imaging. This review summarizes the rapidly evolving field of nanoscale and microscale quantum sensing with spin defects in hBN. We introduce basic properties of hBN spin defects, quantum sensing protocols, and recent experimental demonstrations of quantum sensing and imaging with hBN spin defects. We also discuss methods to enhance their sensitivity. Finally, we envision some potential developments and applications of hBN spin defects.
\end{abstract}
\maketitle

\section{Introduction}

Quantum sensing {\color{black} encompasses} a diverse range of devices that utilize their quantum properties to detect various weak signals \cite{degen2017quantum,budker2007optical}. These devices can achieve high accuracy, stability, and resolution {\color{black} surpassing} the performance of classical {\color{black} approaches}. So far, a diverse range of quantum platforms, including atomic, photonic, and solid-state systems \cite{pirandola2018advances,schirhagl2014nitrogen,castelletto2020silicon}, have  demonstrated  promising performance for quantum sensing applications. In particular, solid-state spin defects have broad applications in condensed matter physics, materials science, biology, and other fields \cite{casola2018probing,kolkowitz2015probing,gross2017real,du2017control,fu2014solar,le2013optical,wolfowicz2021quantum}.

Point defects in solids are analogues of atomic systems ``trapped'' in their host materials. Some point defects exhibit unique optical and spin properties, leading to applications in quantum science and technology. To date, spin color centers in diamond and silicon carbide have been {\color{black} extensively} studied  \cite{jelezko2006single,rose2018observation,koehl2011room,widmann2015coherent,zhang2020material}. In particular, diamond nitrogen-vacancy (NV) centers are currently the most widely used solid-state system for quantum sensing applications \cite{schirhagl2014nitrogen, barry2020sensitivity,hong2013nanoscale}. Diamond NV centers have great stability and  sensitivity to multiple physical quantities over a broad range of probing conditions. They also have large contrasts in optically detected magnetic resonance (ODMR) and  long spin coherence time at room temperature \cite{balasubramanian2009ultralong,kennedy2003long,bar2013solid,herbschleb2019ultra}. However, one of the drawbacks of diamond is the difficulty {\color{black} in creating} high-quality NV centers near the diamond surface due to surface charge and spin noises \cite{romach2015spectroscopy,sangtawesin2019origins,seo2016quantum}. It is also {\color{black} challenging} to fabricate diamond nanostructures and integrate them with other materials. These issues severely limit their applications when close proximity to the target sample is required. Such limitations originate from the dangling bonds on the surface of a three-dimensional (3D) diamond and cannot be fully avoided with diamond. 

\begin{figure*}[htb]
	\centering
	{\resizebox*{14.2cm}{!}{\includegraphics{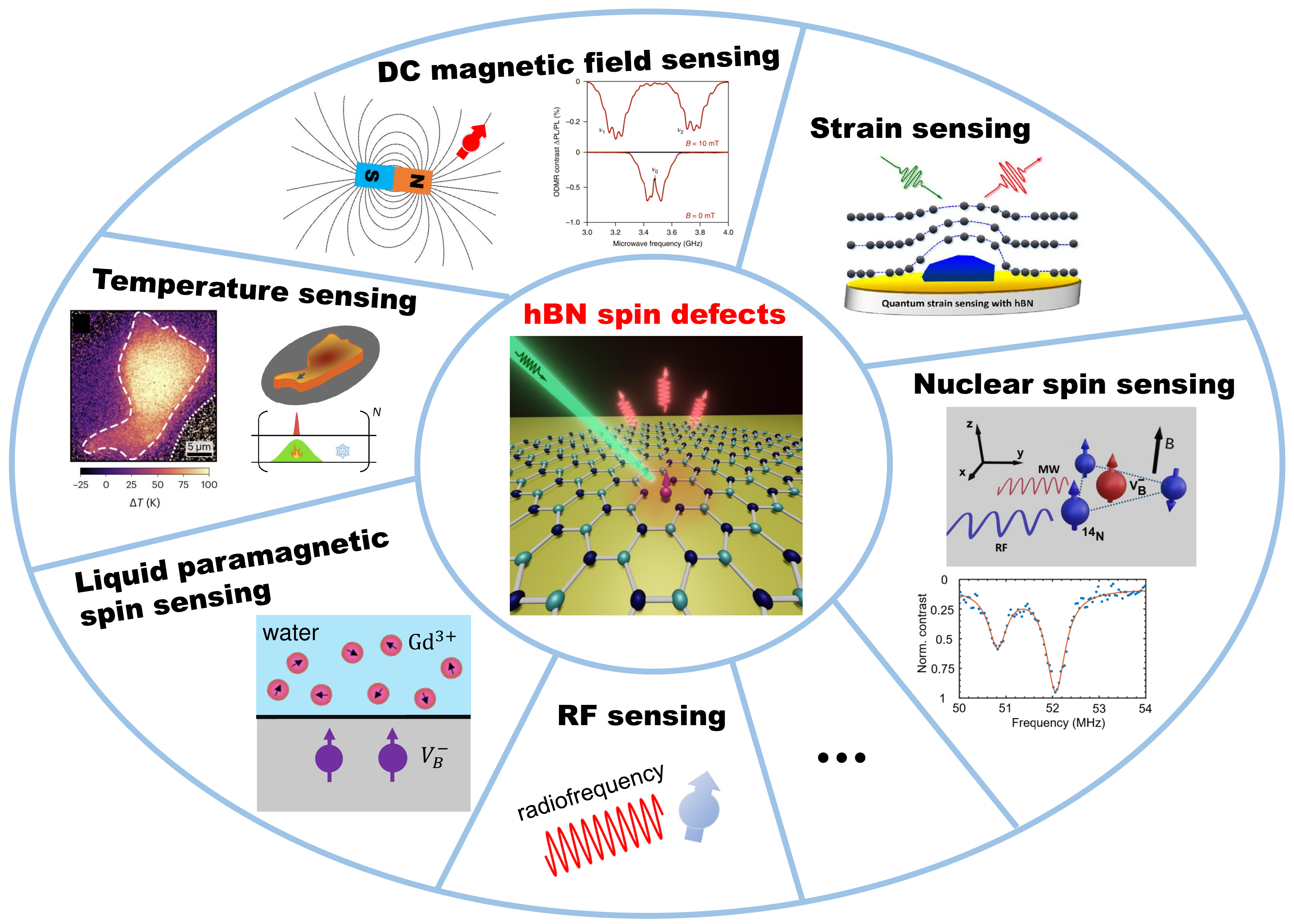}}}
	\caption{An overview of quantum sensing with spin defects in hBN. $V_B^-$spin defects in hBN have been used to measure the DC magnetic field \cite{gottscholl2020initialization,gao2021high,healey2022quantum,huang2022wide}, temperature \cite{gottscholl2021spin,liu2021temperature}, strain \cite{yang2022spin,lyu2022strain}, {\color{black} paramagnetic spins in liquids \cite{robertson2023detection,gao2023quantum}}, nuclear spins \cite{gao2022nuclear}, RF signals \cite{rizzato2022extending}, and etc.  This figure is created from results reported in  \cite{gottscholl2020initialization,healey2022quantum,gao2022nuclear,lyu2022strain,gao2023quantum}. Copyright by Springer Nature \cite{gottscholl2020initialization,healey2022quantum,gao2022nuclear} and American Chemical Society \cite{lyu2022strain}. The center illustration of an hBN spin defect is created by Zhujing Xu.} \label{Figure1:overview}
\end{figure*}

In this context,  quantum defects in two-dimensional (2D) van de Waals (vdW) materials offer new opportunities to {\color{black} address} these limitations \cite{azzam2021prospects,ren2019review,caldwell2019photonics,kubanek2022review}. 2D vdW materials can be stable even at the {\color{black} monolayer limit} and essentially have  no dangling bonds on the surface, which effectively avoids the problems encountered by diamond surfaces. Furthermore, 2D vdW materials can form multifunctional heterostructures, which open prospects for in situ quantum sensing \cite{novoselov20162d}. So far, quantum defects have been discovered in two major classes of 2D materials:  hexagonal boron nitride (hBN) \cite{tran2016quantum,castelletto2020hexagonal,sajid2020single,kianinia2022quantum,aharonovich2022quantum} and transition metal dichalcogenides \cite{srivastava2015optically,he2015single,koperski2015single,chakraborty2015voltage}. In particular, hBN has recently emerged as a promising material for nanoscale quantum sensing. A monolayer hBN has a structure similar to that of graphene but consists of {\color{black} alternating} boron and nitrogen atoms.  hBN has a unique combination of a wide band gap {\color{black} ($\approx$ 6 eV) \cite{cassabois2016hexagonal}}, high thermal conductivity, and excellent chemical and mechanical stability, making it well-suited for use in various sensing conditions where traditional sensing methods may be {\color{black} unreliable} \cite{cassabois2016hexagonal, kianinia2017robust,jungwirth2016temperature,xue2018anomalous}. The large band gap of hBN enables numerous defect energy states within the band gap,  which give rise to many different optically active defect centers with optical spectra {\color{black} ranging} from near-infrared to ultraviolet \cite{tran2016quantum,vogl2019atomic,sajid2020single,abdi2018color,bourrellier2016bright,chejanovsky2016structural,
grosso2017tunable}.

 In 2020, ODMR of spin defects in hBN was reported \cite{gottscholl2020initialization}.  Since then, several types of optically active spin defects in hBN have been {\color{black} observed}  \cite{gottscholl2020initialization,mendelson2021identifying,chejanovsky2021single,stern2022room,ivady2018first}, with the negatively charged boron vacancy ($V_B^-$) spin defect {\color{black} being the most extensively studied}. Many recent studies demonstrated its promising potential as a new platform for quantum sensing \cite{tetienne2021quantum}. The spin states of $V_B^-$ spin defects can be initialized with a green laser, controlled by a microwave,  and read out by their photoluminescence under green laser excitation.

The hBN spin defects can be used to probe external perturbations with high sensitivity \cite{healey2022quantum,gottscholl2021spin}. Their advantage is that these defects are highly stable, residing within a few layers of the material \cite{tran2016quantum,stern2022room}. They provide an opportunity to achieve much closer and stronger  interactions with the target sample to be monitored. In materials science and condensed matter physics, they promise high-resolution imaging  of magnetic fields \cite{song2021direct,fei2018two}, {\color{black} which may} arise from 2D magnetic materials, superconductors, and other spin-based devices. $V_B^-$ spin defects are also sensitive to other physical quantities. Researchers have used hBN $V_B^-$ spin defects for sensing a broad range of quantities (Fig. \ref{Figure1:overview}), including static magnetic fields \cite{healey2022quantum,huang2022wide}, high-frequency magnetic noise due to spin fluctuations in solids \cite{huang2022wide}, temperature \cite{gottscholl2021spin,liu2021temperature}, strain \cite{yang2022spin,lyu2022strain}, nuclear spins \cite{gao2022nuclear}, {\color{black} paramagnetic spins in liquids \cite{robertson2023detection,gao2023quantum}}, and RF signals \cite{rizzato2022extending}.

This review primarily focuses on the $V_B^-$ spin defects in hBN and their applications in sensing (Fig. \ref{Figure1:overview}). For an overview of different spin defects, {\color{black} readers can refer to} a recent review by Liu {\it et al.} \cite{liu2022spin}. 
In the following sections, this paper is organized as follows. We first briefly review the properties of  the $V_B^-$ spin defects and some other spin-active hBN defects in section 2. In Section 3, we discuss basic principles of quantum sensing using hBN spin defects, including the spin Hamiltonian, the sensing protocols, and the sensitivity of this system. Section 4 covers the applications of hBN spin defects in magnetic imaging, temperature sensing, strain sensing, nuclear magnetic resonance (NMR) detection, {\color{black} liquid paramagnetic spin sensing,} and RF sensing. We then discuss ways to improve sensitivity in Section 5 and conclude in Section 6 by envisioning future developments in quantum sensing with hBN spin defects.

\section{Spin defects in hBN}
Optically addressable spin defects offer an interface between photons and electron spin states, which are highly sought after for quantum information applications. In hBN, paramagnetic defects were first measured using conventional electron paramagnetic resonance (EPR)  in 1970s \cite{moore1972electron,katzir1975point,andrei1976point}. Only in the last few years, optically addressable spin defects were observed in hBN \cite{exarhos2019magnetic,gottscholl2020initialization,mendelson2021identifying,chejanovsky2021single,stern2022room}. In this section, we discuss the basic properties of spin defects in hBN and techniques for creating the desired spin defects. 

\subsection {Creation of spin defects}
\begin{figure}[htb]
	\centering
	{\resizebox*{8.5cm}{!}{\includegraphics{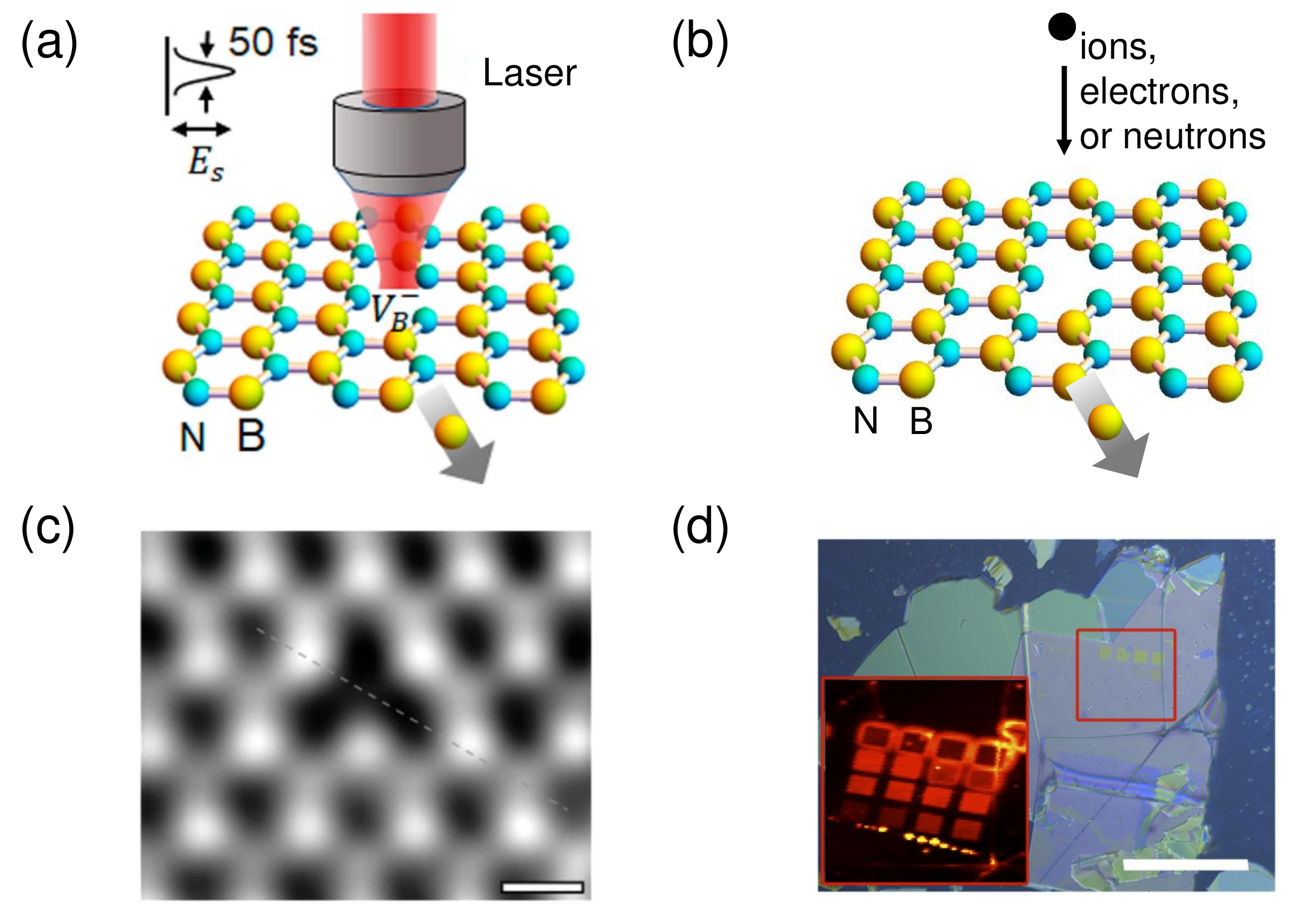}}}
	\caption{ (a) Creation of $V_B^-$ spin defects via femtosecond laser writing. Reproduced from \cite{gao2021femtosecond}.  Copyright of  American Chemical Society. (b) Creation of $V_B^-$ spin defects via ion, electron or neutron irradiation. (c) A TEM image of a boron vacancy defect in an hBN 2D lattice. The scale bar is 0.2 nm. Reproduced from \cite{jin2009fabrication}. Copyright by  American Physical Society. (d) An optical image of an exfoliated hBN flake after ion implantation with a focused ion beam. The inset shows a PL map of the patterned area. The scale bar is 20 $\mu$m. Reproduced from \cite{kianinia2020generation}. Copyright by  American Chemical Society.} \label{Figure2:Creation}
\end{figure}

Since the discovery of optically addressable spin defects in hBN \cite{gottscholl2020initialization}, several experimental approaches have been developed to deterministically create spin defects.  To generate $V_B^-$ defects, the basic idea is to knock out the boron atoms from the hBN lattice (Fig. \ref{Figure2:Creation}). There are many available methods, including neutron irradiation \cite{toledo2018electron,gottscholl2020initialization,li2021defect,haykal2022decoherence}, ion implantation (including focused ion beam) \cite{kianinia2020generation,gao2021high,guo2022generation,baber2021excited}, electron bean irradiation \cite{murzakhanov2021creation} and femtosecond laser writing \cite{gao2021femtosecond} (Fig. \ref{Figure2:Creation} (a)-(b)). Neutron irradiation was the first method used to create $V_B^-$ \cite{gottscholl2020initialization,li2021defect}. Indeed, $^{10}B$ is among the materials with the highest neutron absorption cross-sections and is commonly used as a neutron absorber. The thermal neutrons have a much higher probability to react with boron atoms than nitrogen atoms, which causes fission and leaves mostly  boron vacancies  (Fig. \ref{Figure2:Creation}(c)) with a high production efficiency \cite{doan2014fabrication}. However, this method requires a nuclear reactor with a high thermal flux density, which is hardly accessible. Femotosecond laser writing is a cheaper and more flexible method as it can be operated in an ambient condition \cite{gao2021femtosecond}. However, the spin defect generation process is associated with crystal damage, which has yet to be addressed. High-energy electron irradiation uses a high-energy ($\sim$2 MeV) electron beam to break the chemical bonds, which is a robust way to create $V_B^-$ defects without greatly damaging the hBN crystals \cite{murzakhanov2021creation}.  Ion implantation is so far the most widely used way for generating $V_B^-$ defects, owing to its convenience and ease of access. Many different species of ions can be used to create $V_B^-$ defects, including H$^+$ \cite{murzakhanov2022generation}, He$^+$ \cite{gao2021high,guo2022generation}, C$^+$ \cite{guo2022generation,baber2021excited}, N$^+$ \cite{kianinia2020generation,guo2022generation}, Ar$^+$ \cite{kianinia2020generation,guo2022generation}, Ga$^+$ \cite{kianinia2020generation}, and Xe$^+$ \cite{kianinia2020generation} ions. The $V_B^-$ defects can be generated at different depth by changing the ion energies. The depth as well as the distribution of generated $V_B^-$ defects can be estimated by Stopping and Range of Ions in Matter (SRIM) simulation \cite{ziegler2010srim}. For example, the average depth of defects generated by a 2.5 keV  He$^+$  ion beam is  about 25 nm \cite{gao2021high}. A focused ion beam can pattern an array of $V_B^-$  defects with controllable dose densities and positions \cite{kianinia2020generation}, as shown in Fig. \ref{Figure2:Creation}(d).  {\color{black} Thermal treatment during or after ion implantation is also demonstrated to help improve the spin properties of $V_B^-$ defects \cite{suzuki2023spin}.  Suzuki {\it et al.} performed ion implantation at high temperatures and demonstrated that the ODMR contrast of the created spin defects increased with an increasing temperature \cite{suzuki2023spin}. Such an improvement saturated at around 500~$^\circ$C.}

Besides $V_B^-$ spin defects, there have been efforts to create other spin defects which can be brighter. So far, there have been a few successful approaches to create those spin defects beyond $V_B^-$ \cite{mendelson2021identifying,chejanovsky2021single,stern2022room}. By incorporating carbon atoms during the growth phase of the hBN crystal during the metal-organic vapor-phase epitaxy (MOVPE) and molecular beam epitaxy (MBE),  carbon-related spin defects can be created\cite{mendelson2021identifying,stern2022room}. Carbon ion implantation or femtosecond laser writing followed by thermal annealing have also been used to create spin defects in hBN \cite{mendelson2021identifying,yang2022laser}. Despite the potential of these single spin qubits, research in this area is still in its early stages. In this review, we will mainly focus on   $V_B^-$ spin defects in hBN.

\subsection{Negatively charged boron vacancy defects}

\subsubsection{Electronic structure}
\begin{figure*}[htb]
	\centering
	{\resizebox*{13.5cm}{!}{\includegraphics{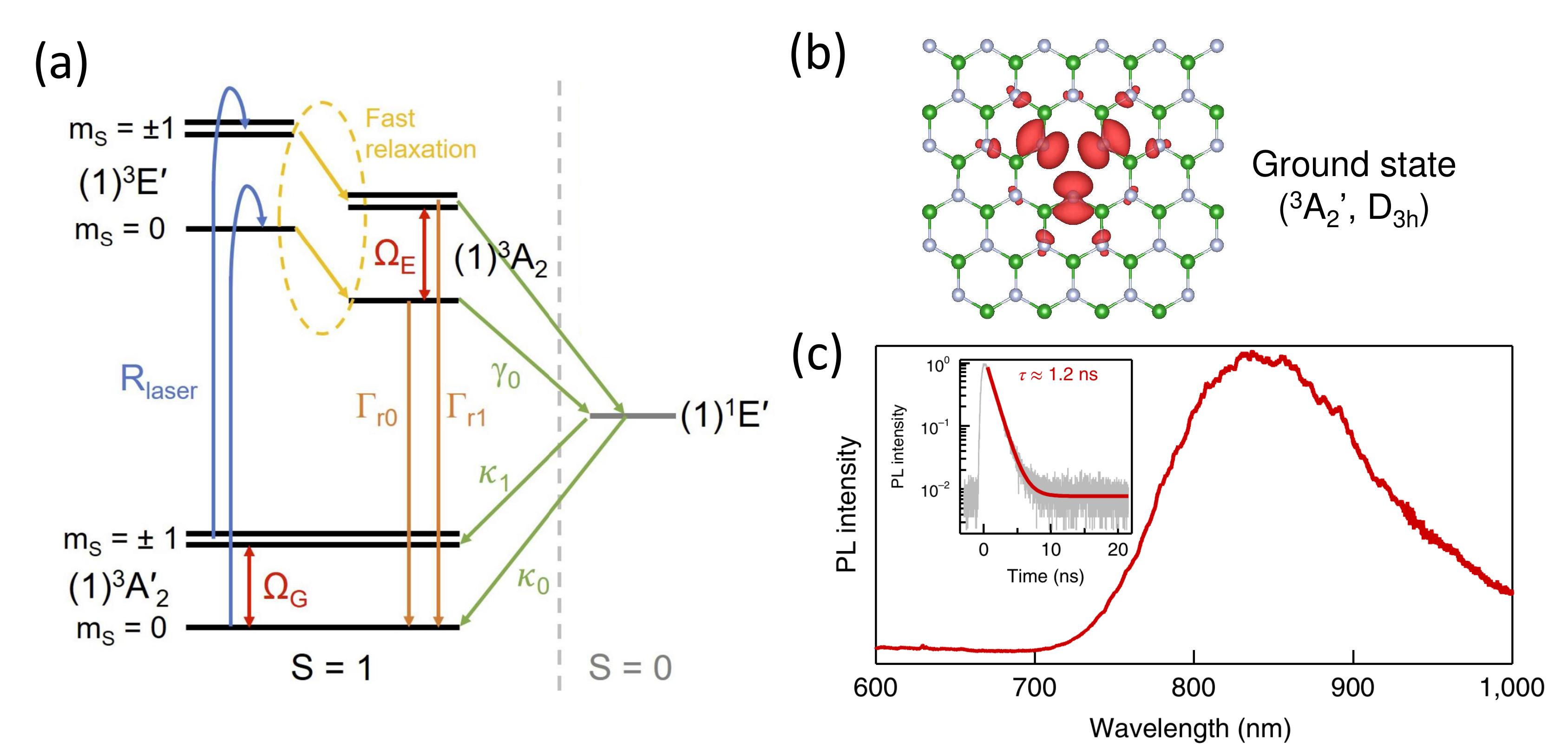}}}
	\caption{(a) An  energy level schematic of a $V_B^-$ defect illustrating the electronic ground state, two excited states and a metastable state. Reproduced from \cite{baber2021excited}. Copyright by  American Chemical Society. (b) Electronic spin density of the $V_B^-$ ground state.  Reproduced from  \cite{gao2022nuclear}. Copyright by  Springer Nature. (c) A photoluminescence spectrum of the $V_B^-$ defect under 532 nm green laser excitation. Reproduced from \cite{gottscholl2020initialization}. Copyright by  Springer Nature. } \label{Figure3:ElectronicStructure}
\end{figure*}

In 2018, Toledo {\it et al.} observed spin defects in a neutron-irradiated hBN sample through conventional EPR experiments \cite{toledo2018electron}. In 2020, Gottscholl {\it et al.} \cite{gottscholl2020initialization} successfully performed ODMR measurements of $V_B^-$ spin defects in hBN. Shortly after the experimental observation, several theoretical studies \cite{reimers2020photoluminescence,sajid2020edge,ivady2020ab,barcza2021dmrg,chen2021photophysical} and excited-state ODMR experiments \cite{mathur2022excited,baber2021excited,mu2022excited,yu2022excited} helped reveal the atomic and electronic structure of the $V_B^-$ defect (Fig. \ref{Figure3:ElectronicStructure}(a)).  A $V_B^-$ defect is formed by a vacant boron site and an additional electron captured from the lattice. It is surrounded by three equivalent nitrogen atoms (Fig. \ref{Figure2:Creation}(c), \ref{Figure3:ElectronicStructure}(b)). The highest point group symmetry of $V_B^-$ is $D_{3h}$, which is expected to be reduced to the  $C_{2v}$ symmetry due to the Jahn-Teller distortion \cite{ivady2020ab} or under the effect of an external strain.  
The electronic structure of the $V_B^-$ defect involves ten electrons. Six are provided by the three fully occupied $p_z$ orbitals of the three nearest nitrogen atoms, and another three are half-filled dangling bonds of the three nearest  nitrogen atoms.  The tenth electron is captured from the environment, making the overall negatively charged state of $V_B^-$. As depicted in Fig. \ref{Figure3:ElectronicStructure}(b), the spin density of the ground state is mostly located in the plane of the hBN lattice, with the highest density near the three nearest  nitrogen atoms.

\subsubsection{Optical properties}
An energy level diagram of the $V_B^-$ defect  is shown in Figure \ref{Figure3:ElectronicStructure} (a). The basic photophysics can be illustrated by three electronic energy levels, including a ground state (GS) of symmetry $^3A_2^\prime$, two excited states (ES) of symmetry $^3E^\prime$ and $^3A_2$, as well as a metastable state with symmetry $^1E^\prime$ (Fig. \ref{Figure3:ElectronicStructure}(a)) \cite{ivady2020ab} . The $V_B^-$ defects can be efficiently excited by light  at most wavelengths below 680 nm \cite{kianinia2020generation}. Experimentally, a 532 nm green laser is commonly used as the excitation light source for  ODMR experiments. At room temperature, $V_B^-$ defects exhibit a broad emission spectrum centered near 820 nm without a sharp distinct zero phonon line (ZPL) \cite{gottscholl2021room} (Fig. \ref{Figure3:ElectronicStructure}(c)). By coupling the hBN layer to a high-Q cavity, Qian {\it et al.} measured the room-temperature ZPL to be at 773 nm \cite{qian2022unveiling}. At cryogenic temperatures, the photoluminescence (PL) spectrum remains mostly unchanged from the room-temperature spectrum \cite{mathur2022excited}. 
The lifetime of the excited state is approximately 1.2 ns, which is mainly due to the fast nonradiative decay \cite{gottscholl2020initialization}. So far, $V_B^-$ defects suffer from low brightness, and no single photon emission has been observed from a $V_B^-$ defect. The low quantum efficiency  ($< 0.1 \%$) of $V_B^-$ defects is because the  transition between the GS and ES triplet subspace is dipole forbidden when strain and electric fields are not considered \cite{ivady2020ab,reimers2020photoluminescence}.

\subsubsection{Spin properties}

\begin{figure*}[bth]
	\centering
	{\resizebox*{12cm}{!}{\includegraphics{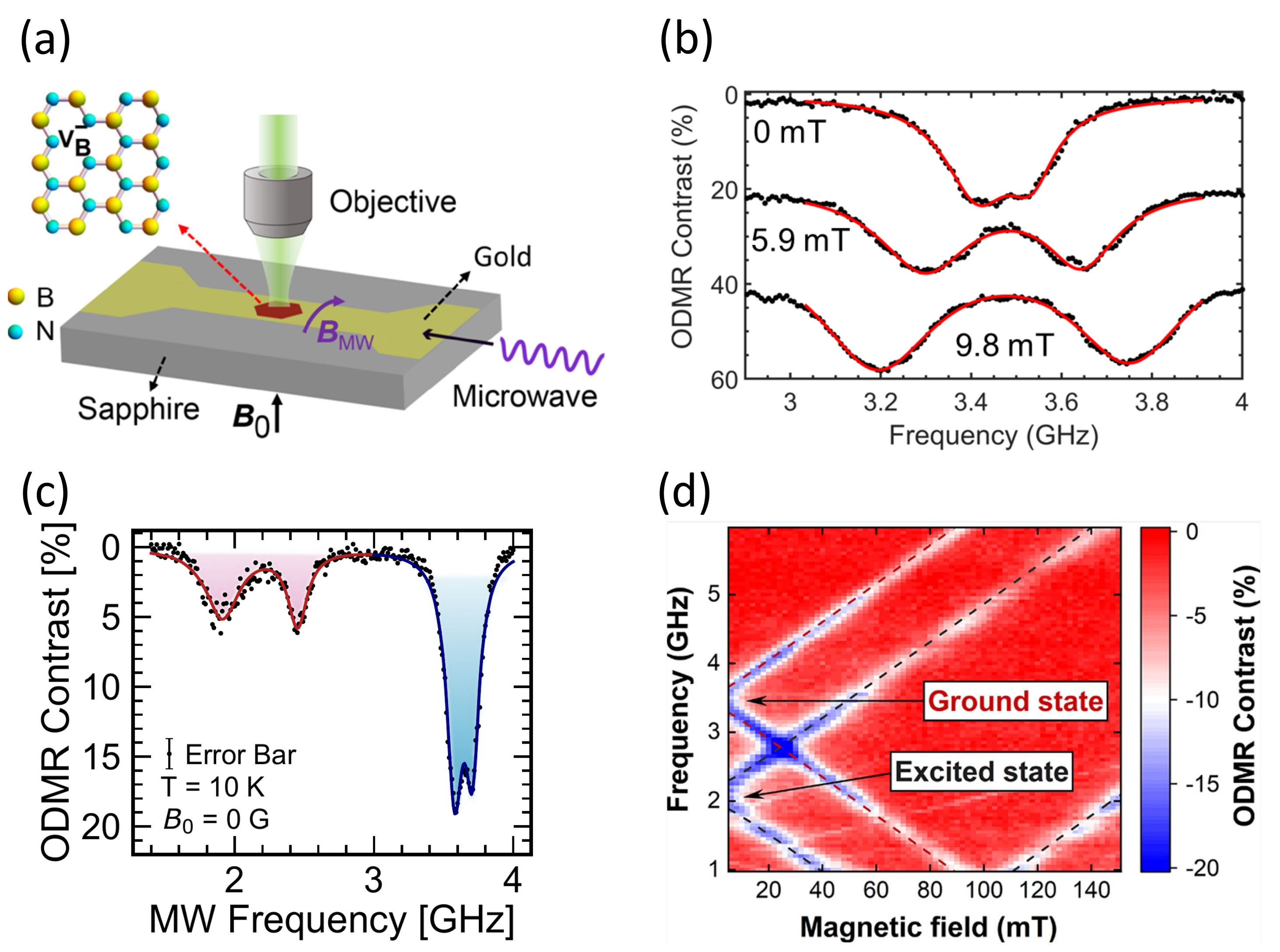}}}
	\caption{(a) An experimental scheme for taking ODMR measurements. (b) ODMR spectra in different magnetic fields: 0 mT, 5.9 mT and 9.8 mT. Reproduced from \cite{gao2021high}. Copyright by  American Chemical Society. (c) The ground state and excited state ODMR spectrum at 0 magnetic field. Reproduced from \cite{mathur2022excited}. (d) The ground state and excited state ODMR spectrum as a function of magnetic field. Reproduced from \cite{baber2021excited}. Copyright by  American Chemical Society.} \label{Figure4:ODMR}
\end{figure*}

The electronic ground state of a $V_B^-$ defect has spin $S=1$. The spin polarization of $V_B^-$ defects is along the hBN c-axis, that is perpendicular to the 2D-lattice plane \cite{gottscholl2020initialization,ivady2020ab,reimers2020photoluminescence}. As shown in Fig. \ref{Figure3:ElectronicStructure}(a), the triplet GS and ES can be further split into three spin sub-levels each. Without an external magnetic field, the dipolar spin-spin interaction gives rise to the fine structures in both GS and ES with longitudinal zero field splitting (ZFS) parameters $D_{gs}$ $\approx$ 3.48 GHz \cite{gottscholl2020initialization} and $D_{es}$ $\approx$ 2.1 GHz \cite{mathur2022excited,baber2021excited,mu2022excited,yu2022excited}. {\color{black} Due to the local electric field or strain, the ODMR spectrum also includes transverse ZFS components.  The ground state transverse ZFS, $E_{GS}$, ranges from  48 MHz to 75 MHz \cite{gottscholl2020initialization,mathur2022excited,gong2022coherent,guo2022generation}, while the excited state transverse ZFS, $E_{ES}$, is reported in the range of 93-154 MHz \cite{mathur2022excited,yu2022excited,baber2021excited,mu2022excited},  making the $\ket{m_s=\pm1}$ states slightly non-degenerate \cite{mathur2022excited}. Differences in  transverse ZFS components among different samples are due to varying local electric fields, which depend on crystal damage levels and charged defect environments. The latter depends on the density of $V_B^-$ ensembles. Both factors are affected by defect creation processes, including doping sources and dosage \cite{guo2022generation}.  } In an external magnetic field, the $\ket{m_s=\pm1}$ states are shifted towards opposite directions via Zeeman effect. The spin lifetime $T_1$ in the ground state  is about 18 $\mu$s at room temperature, and can increase to about 12.5 ms at 20 K \citep{gottscholl2021room}. The spin dephasing time $T_2^*$ is about 100 ns, while the measured spin coherence time $T_2$ varies from about 100 ns \cite{haykal2022decoherence} to about 2 $\mu$s at room temperature \cite{gottscholl2021room}.
There is also a metastable state with $S=0$ which facilitates nonradiative transfer of the spin state from the excited to the ground state. The metastable state is essential for spin polarization and readout.

\subsubsection{Optically detected magnetic resonance}

To perform ODMR measurements, the spin states are initialized and readout optically using a green laser. The optical cycle in the electronic energy level subspace includes multiple steps (Fig. \ref{Figure3:ElectronicStructure}(a)). Under the illumination of a 532 nm laser, the transitions consist of excitation from the GS $^3A_2^\prime$ to the ES $^3E^\prime$, fast (subpicosecond) relaxation to the ES $^3A_2$, radiative recombination back to the GS $^3A_2^\prime$, and nonradiative spin-dependent inter-system crossing (ISC). The spin-dependent ISC process induces the difference in the populations and the PL count rates of different spin states, which is crucial for optical initialization and readout of the spin states (bright state $\ket{m_s=0}$ versus dark states $\ket{m_s=\pm1}$). The laser excitation and radiative decay channels are spin conservative processes, during which the spin state remains unchanged. Due to different MS $\leftrightarrow$ GS/ES transition rates between the $\ket{m_s=0}$ and $\ket{m_s=\pm1}$ states, the defects can be spin polarized to the $\ket{m_s=0}$ state using continuous optical pumping. This enables us to initialize and readout the spin states and perform quantum sensing based on various protocols as described in later sections.

Fig. \ref{Figure4:ODMR}(a) shows the schematic of an ODMR measurement setup involving optical excitation, and microwave drive using a microwave waveguide onto which the hBN flakes are transferred \cite{gao2021high}. The objective lens helps focus the excitation laser and collect the PL emission from the defects. An external magnetic field can be applied with a nearby magnet.

\begin{figure*}[htb]
	\centering
	{\resizebox*{14.2cm}{!}{\includegraphics{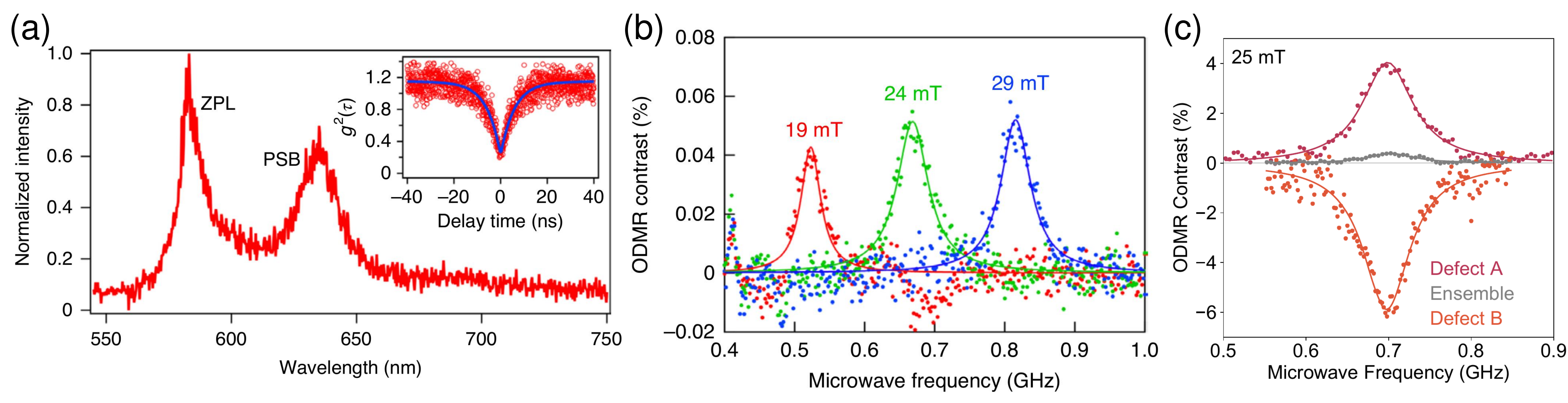}}}
	\caption{(a) A photoluminescence spectrum of a carbon-related defect in hBN. Inset: $g^2(\tau)$ measurement demonstrating single-photon emission from a single spin defect. Reproduced from \cite{mendelson2021identifying}. Copyright by  Springer Nature. (b) ODMR spectra of a carbon-related defect, which demonstrates a shift in the peak as a function of the magnetic field. Reproduced from \cite{mendelson2021identifying}. Copyright by  Springer Nature. (c) ODMR spectra of two single spin defects and an ensemble. The ODMR spectra of single spin defects can have opposite signs, leading to an overall reduction in the ODMR contrast of an ensemble with different types of spin defects.  Reproduced from \cite{stern2022room}.} \label{Figure5:otherDefects}
\end{figure*}

A basic ODMR experiment is to record the electron spin resonance (ESR) spectrum of the $V_B^-$ defects by slowly sweeping the driving microwave frequency over the ESR transitions. A reduction of the PL intensity will be observed when the microwave frequency matches the ESR transition frequency so that the transition between $\ket{m_s=0}$ and $\ket{m_s=\pm1}$ states occurs (Fig. \ref{Figure4:ODMR}(b)). When spin defects are in the presence of an external magnetic field, the transition frequencies shift due to the Zeeman effect (Fig. \ref{Figure4:ODMR}(b)) \cite{gottscholl2020initialization,gao2021high}. The ODMR experiments can be carried out using a continuous wave (CW) protocol or pulsed microwave measurements. The CW ODMR protocol is widely used in sensing applications as it is technically simple and offers relatively high photon counts.  When we sweep the microwave frequency over a large range, both the ground-state and the excited-state transitions are clearly visible, forming two distinct sets of branches in the ODMR spectrum as shown in Fig. \ref{Figure4:ODMR}(c). By fitting the slope of the transition frequencies versus magnetic fields (Fig. \ref{Figure4:ODMR}(d)), the electron spin Land{\'{e}} g-factor of $V_B^-$ can be obtained as g $\approx$ 2, indicating the  spin-orbital coupling is very weak \cite{gottscholl2020initialization,mathur2022excited}. ODMR measurements with low microwave powers can also reveal the hyperfine coupling between the electronic spin and the surrounding nitrogen nuclear spins. This coupling results in seven hyperfine peaks in the GS ODMR spectrum, which will be discussed more in Sec. 4.5 \cite{gottscholl2021room, gao2022nuclear}. The hyperfine splittings have not been resolved in the ES due to the short ES lifetime and inhomogenous broadening in the defect ensemble \cite{mathur2022excited,baber2021excited,mu2022excited,yu2022excited}.

\subsection {Other hBN spin defects}

While the $V_B^- $ defect is the most studied hBN spin defect so far, other spin defects in hBN have been demonstrated experimentally. Carbon was identified as a source of a family of defects, and shown to have ODMR signals (Fig. \ref{Figure5:otherDefects})  \cite{mendelson2021identifying, stern2022room}. By controlling the concentration of carbon impurities, both spin ensembles (with high concentration) (Fig. \ref{Figure5:otherDefects}(a)) and single spin defects (with low concentration) can be generated. Mendelson {\it et al.} first generated carbon-related spin defects in hBN by the inclusion of carbon atoms during the  metal–organic vapour-phase epitaxy (MOVPE) growth of hBN samples, and demonstrated the room-temperature ODMR in the spin ensembles \cite{mendelson2021identifying} (Fig. \ref{Figure5:otherDefects}(b)). Stern {\it et al.} later reported ODMR measurements on single spin defects with the carbon-doped hBN sample,  also created by the MOVPE method \cite{stern2022room}. The ODMR signal was found to change in sign for different single spin defects, even within the same sample (Fig. \ref{Figure5:otherDefects}(c)). Chejanovsky {\it et al.} identified a few types of spin defects in hBN flakes, which had been annealed at $850^{\circ}$C \cite{chejanovsky2021single}. The ODMR spectra were observed only at cryogenic temperature in \cite{chejanovsky2021single}. One of those defects has been identified to be a carbon substitution defect \cite{auburger2021towards}. Recently, Guo {\it et al.} reported a room-temperature ultrabright single spin defect with a ZPL around 540 nm \cite{guo2021coherent}.  This spin defect exhibits $\approx$ 2.5 MHz count rate under 1 mW green laser excitation, which is much brighter than other reported single spin defects in hBN. All of the hBN single spin defects discussed above exhibit g-factors $\approx$2, and have small ZFS ($<$ 100 MHz) which often requires an external magnetic field to obtain a clear ODMR signal.

In addition, Chen and Quek theoretically predicted a neutrally-charged double-boron-vacancy (VB2) spin defect with a triplet GS \cite{chen2021photophysical}. The ZFS parameters were proposed to be $D_{gs} = -1.1$ GHz and $E_{gs} = -76.8$ MHz. Babar {\it et al.} experimentally observed this defect via scanning transmission electron microscopy (STEM) \cite{babar2021quantum}. A VB2 defect consists of a pair of adjacent in-plane boron vacancies and a nitrogen atom migrating into one of the vacancy sites. As a result, this forms a V$_B$V$_N$N$_B$ complex including a boron vacancy, a nitrogen vacancy, and a nitrogen anti-site. An oxygen related defect has also been identified via conventional EPR and PL spectra simulations and experiments \cite{li2022identification}. In addition, {\color{black} Yang} {\it et al.} reported new types of spin defects in hBN created by femtosecond laser writing and subsequent annealing, but the origins of these defects are still unknown \cite{yang2022laser}. A theoretical calculation by Pinilla {\it et al.} also predicted that non-adjacent carbon substitutions of the kind $C_N$-$C_N$ or $C_B$-$C_B$ have triplet ground state which may likely be spin defects \cite{pinilla2022carbon}.

\begin{table*}[htb]
	\caption{Coupling coefficients and  \textcolor{black}{reported sensitivity}}
	\begin{ruledtabular}
		{\begin{tabular}{lcccccc} 
				Property & Coupling Coefficient  & \textcolor{black}{Reported Sensitivity (295 K)} & \textcolor{black}{Sensor Volume} & Reference  \\ 
				Magnetic field & $\gamma_e $ = 28.0 kHz/$\mu$T & 85.1 $\mathrm{ \mu T/\sqrt{Hz}}$ & $\sim$100 $\mu$m$^3$ & \cite{gottscholl2021spin}  \\ 
				&    &  73.6 $\mathrm{ \mu T/\sqrt{Hz}}$  & $\sim$0.0007 $\mu$m$^3$ & \cite{sasaki2023magnetic}  \\ 
				&    &  8 $\mathrm{ \mu T/\sqrt{Hz}}$  & $\sim$0.007 $\mu$m$^3$ & \cite{gao2021high}  \\ 
				&    &  2.55 $\mathrm{ \mu T/\sqrt{Hz}}$  & $\sim$0.02 $\mu$m$^3$ & \cite{liang2022high}  \\ 
				In-plane strain & $\partial D/ \partial \eta_a$ = -81 GHz & $10^{-5}$ &$ \sim$0.002 $\mu$m$^3$  & \cite{yang2022spin}  \\ 
				Out-of-plane strain & $\partial D/ \partial \eta_c$ = $ $-24.5 GHz$ $ &  $10^{-4}$& $\sim$0.005 $\mu$m$^3$ & \cite{lyu2022strain}  \\ 
				Temperature & $\partial D/ \partial T$ = -623    kHz/K & 3.82  $\mathrm{K/\sqrt{Hz}}$& $\sim$100 $\mu$m$^3$ & \cite{gottscholl2021spin}  \\ 
				X Pressure & $\partial D/ \partial P_x$ = -0.136  Hz/Pa & 17.5 $\times 10^6$ $ \mathrm{ Pa/\sqrt{Hz}}$ & $\sim$100 $\mu$m$^3$& \cite{gottscholl2021spin}  \\ 
				Y Pressure & $\partial D/ \partial P_y$ = -0.212 Hz/Pa & 11.2 $\times 10^6$ $ \mathrm{ Pa/\sqrt{Hz}}$ & $\sim$100 $\mu$m$^3$& \cite{gottscholl2021spin}  \\ 
				Z Pressure & $\partial D/ \partial P_z$ =  -0.91 Hz/Pa & 2.62 $\times 10^6$ $ \mathrm{ Pa/\sqrt{Hz}}$ &$\sim$100 $\mu$m$^3$ & \cite{gottscholl2021spin}  \\ 
		\end{tabular}}
	\end{ruledtabular}
	\label{sample-table}
\end{table*}

\section{Quantum sensing with hBN spin defects}
Spin defects in hBN show improved versatility for quantum sensing of proximate objects, e.g., 2D magnetic materials. In this section, we introduce the basic principles of quantum sensing using hBN spin defects and related sensing protocols.

\subsection {Spin Hamiltonian}

The ground-state  electron spin Hamiltonian $H_{gs}$ of $V_B^- $ involves terms with ZFS, electron Zeeman splitting,  and electron-nuclear spin hyperfine interaction  \cite{gottscholl2020initialization,gao2022nuclear,gracheva2023symmetry}:
\begin{equation}
\begin{split}
		H_{gs}  = & D_{gs}[S_z^2-S(S+1)/3]+E_{gs}(S_x^2-S_y^2)  \\
		                & +\gamma_{e} {\bf B} \cdot {\bf S} + \sum_{k=1,2,3} {\bf S} {\bf A}_k {\bf I}_k, 
\end{split}\label{Hamiltonian:eq1}
\end{equation}
where $D_{gs} \approx 3.48$ GHz is the ground state  longitudinal ZFS parameter \cite{gottscholl2020initialization}, $E_{gs} \approx 48$ MHz is the transverse ZFS parameter \cite{mathur2022excited},  $\bf{S}$ and $S_j$ (j=x, y, z) are electron spin-1 operators,  $\gamma_{e}=$28 GHz T$^{-1}$  
 is the electron-spin gyromagnetic ratio when the Land{\'{e}} g-factor is $g=2$.   ${\bf I}_k $  (k=1, 2, 3) is the nuclear spin-1 operator of the three nearest  $^{14}N$ nuclei, and $\bf{A_{k}}$ is the hyperfine interaction tensor.  An important component of the hyperfine interaction tensor is $A_{zz} \approx 48$ MHz \cite{ivady2020ab,gracheva2023symmetry,gao2022nuclear}.

 When the external magnetic field ${B_0}$ is parallel to the $c$ axis, 
 the resonant frequencies of transitions between the $\ket{m_s=0}$ and $\ket{m_s=\pm1}$ states are \cite{gottscholl2020initialization}:
\begin{equation}
	\nu_{\pm}=D_{gs} \pm\sqrt{E_{gs}^2+ (\gamma_{e} B_0)^2}. \label{ODMRFreq:eq2}
\end{equation}
The formula for the excited state is similar, but has different parameters  \cite{mathur2022excited}. The experimentally observed ODMR spectrum of $V_B^-$  is shown in Fig. \ref{Figure4:ODMR}(d). The slope of the lines correspond to the electron spin gyromagnetic ratio.

\begin{figure*}[htb]
	\centering
	{\resizebox*{17cm}{!}{\includegraphics{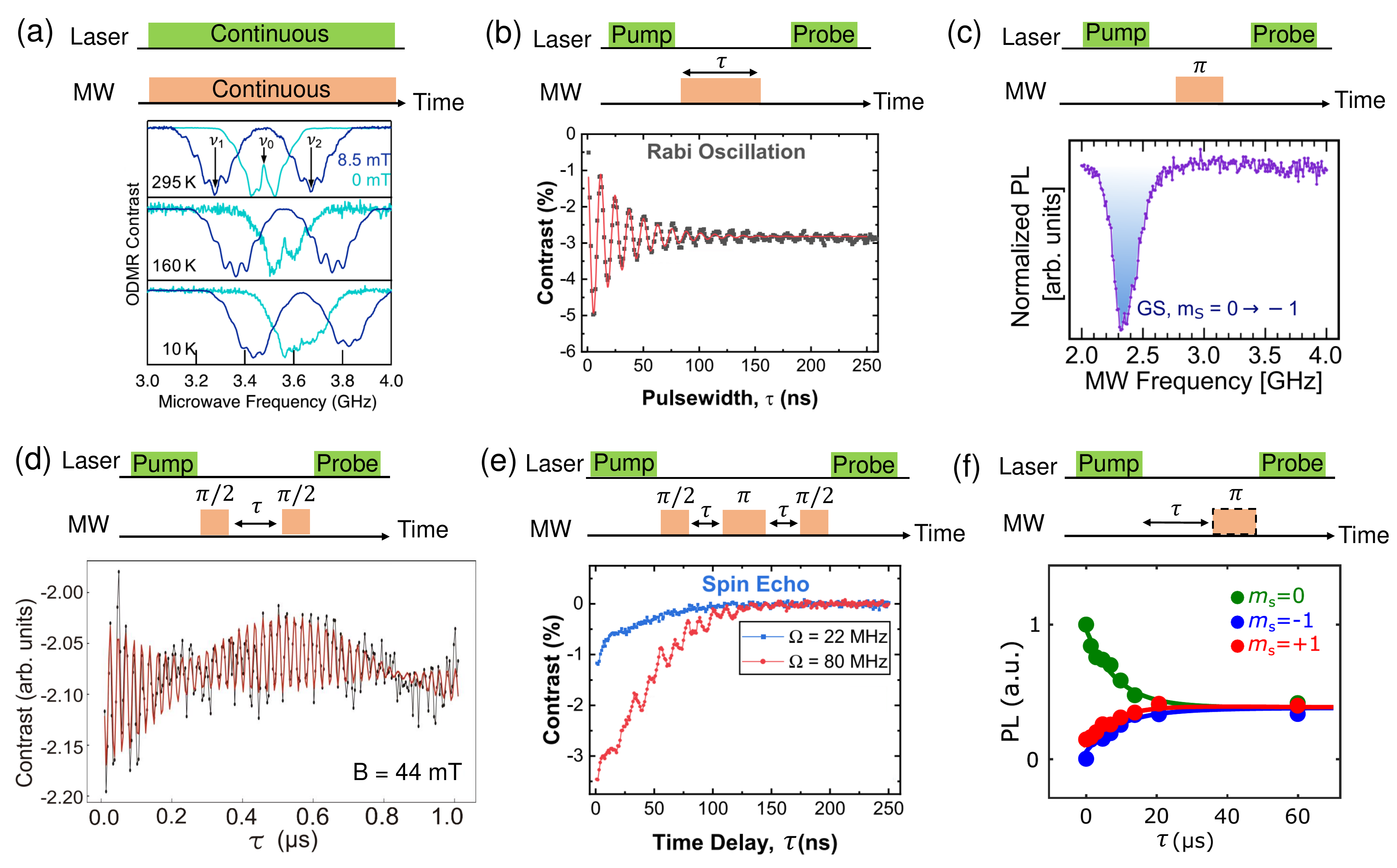}}}
	\caption{(a) Top: Illustration of CW ODMR pulse sequence. Bottom: CW ODMR Spectra of $V_B^-$ spin defects as a function of temperature and magnetic field \cite{gottscholl2021spin}. (b) Top: Pulse schematic for the coherent control experiment. A microwave pulse with a variable time duration time is applied between two laser pulses to manipulate the spin state. Bottom: An example of Rabi oscillation  \cite{ramsay2023coherence}. (c) Top: Pulse schematic for the pulsed ODMR measurement. Bottom: An example of pulsed ODMR measurement in a 400 G magnetic field \cite{mathur2022excited}.   (d) Top: Pulse schematic for the Ramsey experiment. Bottom: An example result of Ramsey fringes of $V_B^-$ defects due to surrounding nuclear spins \cite{liu2022coherent}. (e) Top: Pulse schematic for the spin echo measurement. Bottom: Spin echo measured by using two different microwave driving powers \cite{ramsay2023coherence}. With a higher microwave Rabi frequency, $V_B^-$ shows a better spin coherence. (f) Top: Pulse schematic for relaxometry measurements. The dashed box refers to the application of microwaves during the measurement of the states $m_s=\pm1$ and is absent during the measurement of $m_s=0$. Bottom: Normalized photoluminescence of the three spin sublevels as a function of delay time \cite{huang2022wide}. Reproduced from \cite{gottscholl2021spin, huang2022wide, ramsay2023coherence, mathur2022excited, liu2022coherent}.} \label{Figure6:Protocols}
\end{figure*}

Quantum sensing can be performed to measure  magnetic field, temperature, strain and pressure via the shift in the ODMR resonance frequency \cite{gottscholl2021spin}. As seen in Fig \ref{Figure4:ODMR}(b), the separation between the two dips in the ODMR spectrum increases  when the external magnetic field increases \cite{gottscholl2020initialization,gao2021high}. This is due to the different signs of the Zeeman interaction for $\ket{m_s=+1}$ and $\ket{m_s=-1}$ states. The two peaks have Zeeman shifts in opposite directions, yielding a separation of  $2\gamma_e B_0$ when $\gamma_e B_0 \gg E_{gs}$. This enables accurate measurement of  magnetic fields. 
The center of the GS ODMR transition frequencies ($D_{gs}$) is proved to be temperature dependent as shown in Fig \ref{Figure6:Protocols}(a), which allows temperature sensing based on ODMR experiments \cite{gottscholl2021spin,liu2021temperature}. The pressure also affects hBN $V_B^-$ defects via the ZFS parameter $D_{gs}$. By measuring the change of $D_{gs}$, one can obtain the ambient pressure with $V_B^-$ defects. The coupling constants and reported sensitivity of recent experiments at room temperature are summarized in Table 1. {\color{black} The sensor volume in Ref. \cite{gottscholl2021spin} is relatively large as it used a low NA objective lens and a thick hBN single crystal with low-density spin defects. It can be substantially reduced by using a high NA objective lens and thin hBN nanosheets.} For strain sensing, estimated resolutions based on figures reported in \cite{yang2022spin,lyu2022strain} are listed here since strain sensing sensitivities have not been reported. More details of these experiments will be discussed in Section 4.

\subsection{Sensing protocols}

In order to measure external fields via shifts in the ODMR signal, a number of protocols \cite{degen2017quantum,rondin2014magnetometry} have been developed and most of them are applicable to the hBN $V_B^-$ spin defects. The most common protocols are the CW ODMR spectroscopy, pulsed time evolution measurement, and spin relaxometry measurement. Here we briefly explain each of the three methods.

\subsubsection{Continuous wave optically detected magnetic resonance} 
In the CW ODMR protocol, the resonant frequencies of the spectrum are recorded and the shifts are tracked in the presence of external fields (Fig. \ref{Figure6:Protocols}(a)) \cite{gottscholl2020initialization,gottscholl2021spin}. These shifts contain the information of the magnitudes of the external perturbations as per the Hamiltonian discussed in Equation \ref{Hamiltonian:eq1}. This method does not require pulsed laser excitation, precise microwave manipulation, fast photodetectors, or multichannel pulse generators. Therefore, this method is easy to implement for real sensing applications. Fig \ref{Figure6:Protocols}(a) shows  CW ODMR spectra of hBN $V_B^-$ defects as a function of magnetic field and temperature   \cite{gottscholl2021spin}. The resonant frequency shifts with external perturbations, from which the magnitude of the magnetic field or temperature can be calculated. 

The smallest possible detectable shift in the resonant frequency of ODMR is \cite{rondin2014magnetometry}
\begin{equation}
	\Delta \omega\approx\frac{\Delta \nu}{C\sqrt{I_0T}},
\end{equation}
where $\Delta \nu$ is the linewidth of the spectrum, $C$ is the optical contrast between the $\ket{0}$ and $\ket{\pm1}$ states, $I_0$ is the detected rate of photons (photon counts per second), and $T$ is the collection time per measurement. This gives a static magnetic field detection sensitivity $\eta$ \cite{rondin2014magnetometry}:
\begin{equation}
	\eta \approx \frac{\Delta \nu}{\gamma_e C\sqrt{I_0}}.
\end{equation}
The sensitivity can be improved by increasing the collected photon count rate and ODMR contrast, as well as narrowing down the ODMR linewidth. These parameters are dependent on laser and microwave powers, and can change oppositely when we tune the laser or microwave powers. Thus there is a  trade-off when we optimize the sensitivity. {\color{black} The sensitivity can also be improved by using a larger number of defects by increasing the density of spin defects or increasing the sensor volume. However, increasing the spin density too much could increase the linewidth due to spin-spin interaction. Increasing the sensor volume will reduce the spatial resolution.} In addition, spin ensembles usually suffer from inhomogeneous broadening, which may need to be considered for optimizing the ultimate sensitivity of a spin ensemble. 

\subsubsection{Pulsed measurements}
Pulsed measurement is an alternative method that can be more sensitive to certain external perturbations. Pulsed sensing protocols rely on microwave pulses when the laser is turned off during the sensing period.  This technique can avoid optical and microwave power broadening, and enable nearly $T_2^*$-limited sensitivities in pulsed ODMR and Ramsey measurements \cite{taylor2008high,barry2020sensitivity}. By using more complex pulsed sensing protocols, like spin-echo and dynamic decoupling, the sensitivity can be further improved to be limited by $T_2$ instead of $T_2^*$ \cite{taylor2008high,barry2020sensitivity}. 
Three major groups of pulsed sensing protocols are: pulsed-ODMR, Ramsey and spin-echo. All these techniques require precise coherent control of spins. 

a) Spin coherent control

Coherent control of $V_B^-$ spin defects is achieved by applying a microwave pulse of variable length to coherently manipulate the spin state (Fig. \ref{Figure6:Protocols}(b)) \cite{gottscholl2021room,ramsay2023coherence,fuchs2008excited}. Depending on the frequency of the microwave pulse, we can efficiently couple two spin states, e.g. $\{\ket{m_s=0},\ket{m_s=-1}\}$. After initializing the spin state to $\ket{m_s=0}$ state with a green laser pulse, a periodically varying magnetic field B$_{MW}(t)$ (microwave), with orientation perpendicular to the $V_B^-$ spin axis and frequency $\nu_-$ in resonant with the $\ket{m_s=0}$ $\leftrightarrow$
$\ket{m_s=\-1}$ transition, can be applied for coherent control. The microwave magnetic field drives the spin  to oscillate between the $\ket{m_s=0}$ state and the $\ket{m_s=-1}$ state at an angular frequency $\Omega_R\propto$ B$_{MW} $, which is the Rabi frequency.  This oscillation is also known as the Rabi oscillation. After the microwave pulse, another laser pulse is applied and meanwhile a  photon counter  records the photon numbers from the spin defects. Since the  $\ket{m_s=0}$ and $\ket{m_s=\-1}$ states have different brightness, the final spin state can be read out based on the collected photon counts. By varying the time duration  of the microwave pulse, an oscillation of the PL intensity is observed, from which we can extract the Rabi frequency that will be useful for other pulsed experiments.

b) Pulsed ODMR measurement

In a pulsed ODMR measurement (Fig. \ref{Figure6:Protocols}(c)), we can obtain a spectrum similar to the ground-state part of the CW ODMR but without suffering from optical and microwave broadening. In the pulsed ODMR sensing protocol \cite{barry2020sensitivity}, the spin defects are first optically initialized to $\ket{m_s=0}$ state. Then a microwave is applied with a finite duration time $\tau_\pi$ = $\pi/\Omega_R$. Such a microwave pulse is called a $\pi$ pulse. Finally, a laser pulse is used to optically readout the final spin state. In the experiment, the microwave frequency is swept across the resonance. At the resonant frequency, the $\pi$ pulse flips the spin state most precisely from $\ket{m_s=0}$ to $\ket{m_s=-1}$, which gives the highest ODMR contrast. A change in the external field will shift the spin resonance away from the original transition frequency, and thus changes the population transferred to the $\ket{m_s=-1}$ state. For the pulsed ODMR, the spectral linewidth is determined by the natural profile of spin transitions  and broadening from the duration of the microwave $\pi$ pulse. Both factors are limited by the spin dephasing time $T_2^*$, which gives a $T_2^*$ limited-sensitivity to DC magnetic fields \cite{dreau2011avoiding}:
\begin{equation}
		\eta_{DC} \approx \frac{{\color{black}1}}{\gamma_e C\sqrt{I_0T_CT_2^*}}.
\end{equation}
where $T_C$ is the photon counting time in each pulse. The contrast $C$ decreases when $T_C$ is too large. Thus there is an optimal $T_C$ to achieve the highest sensitivity.

c) Ramsey measurement

Ramsey sequence is another method that is sensitive to static DC magnetic fields (Fig. \ref{Figure6:Protocols}(d)) \cite{schirhagl2014nitrogen}. Similar to the pulsed ODMR technique, the sensitivity of a Ramsey measurement is limited by $T_2^*$. The Ramsey experiment also starts with laser initialization to prepare the spin in the $\ket{m_s=0}$ state. Then a resonant microwave field $B_{MW}$ is applied for a duration $\pi/(2\Omega_R)$. This is known as a $\pi/2$ pulse, which transfer the initial spin state into a superposition state $(\ket{m_s=0}+\ket{m_s=-1})/\sqrt{2}$. This state is then left to precess for a duration of $\tau$. If the spin defect is in  a DC magnetic field, an additional phase difference $\phi$ will be accumulated between the $\ket{m_s=0}$ state and the $\ket{m_s=-1}$ state. The phase $\phi$ is proportional to the duration time and the magnitude of the DC field. After the precession period, a second $\pi/2$ pulse is applied to map the phase $\phi$ onto a population difference between the $\ket{m_s=0}$ state and the $\ket{m_s=-1}$ state, which will be  readout optically by using a second laser pulse. As depicted in Fig. \ref{Figure6:Protocols}(d), a oscillation will be observed as a function of $\tau$ in the Ramsey measurement. The oscillation period depends on the external magnetic field and the local environment of the spin defect. 

d) Spin echo measurement

The sensitivity of both the pulsed ODMR and the Ramsey measurement is limited by $T_2^*$, which is relatively short for solid-state spin defects \cite{maze2008nanoscale,grinolds2013nanoscale}. For detecting narrow band AC fields, there are multiple dynamical decoupling techniques that can overcome the $T_2^*$ limitation and achieve better sensitivities . Among dynamical decoupling methods, Hahn echo is the simplest one to implement (Fig. \ref{Figure6:Protocols}(e)) \cite{degen2017quantum}. The Hahn echo is built upon the Ramsey sequence with an additional microwave $\pi$ pulse in the middle of the two $\pi/2$ pulses. This additional $\pi$ pulse refocuses the dephased spin states and extends the spin coherence beyond $T_2^*$. In this case, the decay time of the spin coherence is called $T_2$ which is usually much longer than $T_2^*$. However, since the $\pi$ pulse flips the spin state, a DC magnetic field is decoupled from spins. Therefore, such sensing protocols are more sensitive to oscillating magnetic fields.  The magnetic field sensitivity of a spin defect to an oscillating AC magnetic field is: 
\begin{equation}
		\eta_{AC} {\color{black}\approx} \eta_{DC} \sqrt{\frac{T_2^*}{T_2}}.
\end{equation}
 
\subsubsection{Spin relaxometry measurement}

\begin{figure*}[htb]
	\centering
	{\resizebox*{14cm}{!}{\includegraphics{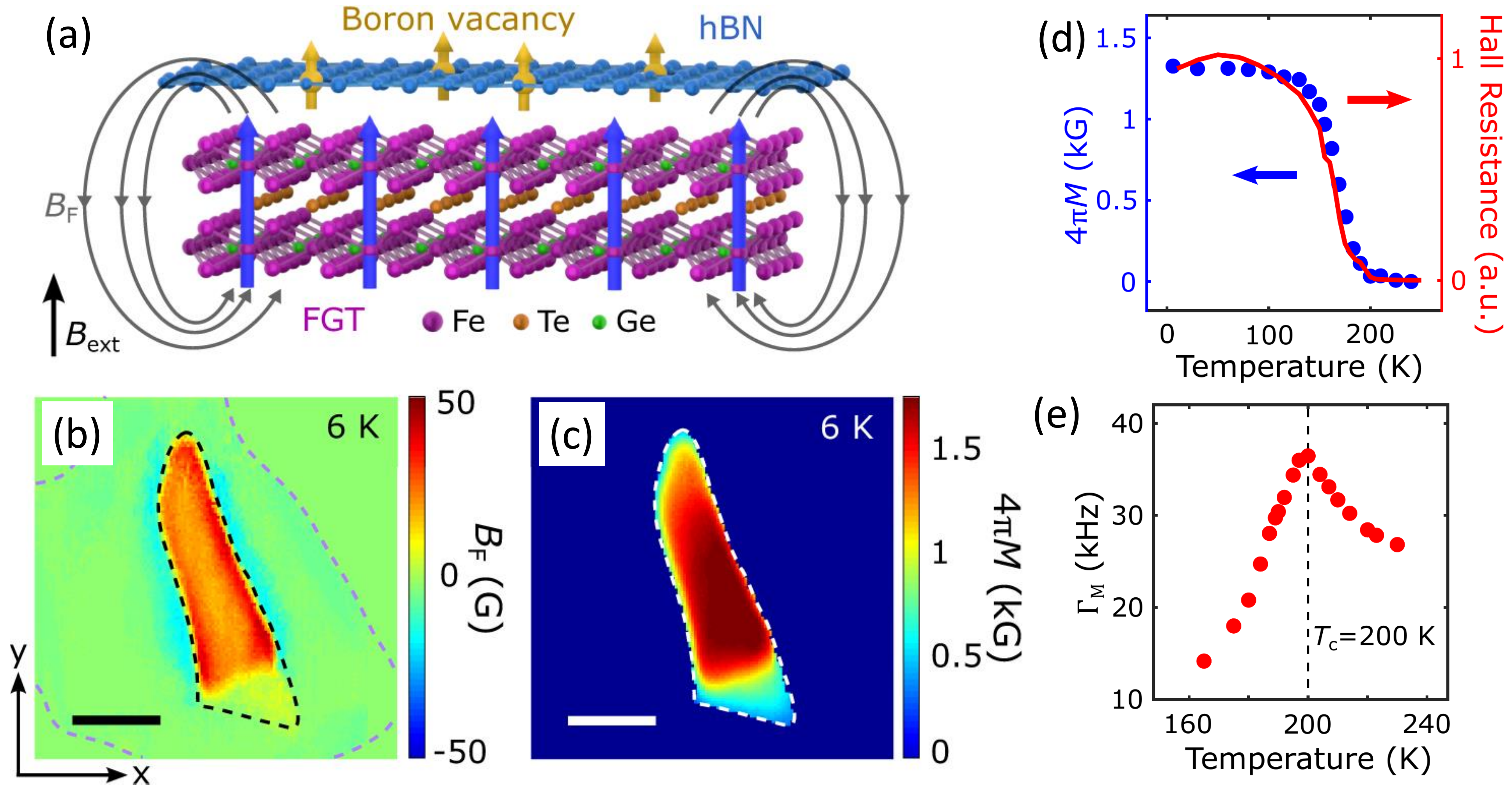}}}\hspace{5pt}
	\caption{(a) Schematic illustration of quantum sensing of local stray fields generated by a 2D ferromagnetic material. (b) A two-dimensional map of the static stray magnetic field from an exfoliated FGT flake at 6 K with an external perpendicular magnetic field of 142 G. {\color{black} The scale bar is 5 $\mu$m.} (c) Reconstructed magnetization 4$\pi M$ of the exfoliated FGT flake. {\color{black} The scale bar is 5 $\mu$m.} (d) Temperature dependence of spatially averaged magnetization of the FGT flake. (e) Change of the relaxation rate $\Gamma_M$ of  $V_B^-$ spin defects near an FGT flake showing a maxima of magnetic noise near the Curie temperature.  Reproduced from \cite{huang2022wide}.} \label{Figure7:MagneticSensing}
\end{figure*}

Spin relaxometry can probe fast fluctuating signals with frequencies from megahertz to gigahertz \cite{steinert2013magnetic,huang2022wide}. This sensing method compares the spin relaxation time $T_1$  of bare spin defects and the spin defects in the presence of the target fluctuating fields. Fig. \ref{Figure6:Protocols}(f) shows the pulse sequence of a spin relaxometry experiment. A faster decay of $V_B^-$  spin states  was observed when hBN spin defects were placed near a 2D magnetic material \cite{huang2022wide}, which was due to the magnetic noise generated by spin fluctuations in the magnetic material.

\section {Emerging applications of hBN spin defects}

As a layered vdW material, hBN enables nanoscale proximity between the hBN spin defects and a target sample, especially another layered 2D material. This offers a new platform for studying the local physical quantities with high precision and spatial resolution. So far, spin defects in hBN have been used for many different sensing applications. In this section, we review recent achievements in this exciting field.

\subsection{Magnetic field and spin noise sensing and imaging}

 Recent studies have integrated hBN with built-in $V_B^-$ spin defects and 2D  ferromagnetic (FM) materials, such as Fe$_3$GeTe$_2$ (FGT) (Fig \ref{Figure7:MagneticSensing}) \cite{huang2022wide} and CrTe$_2$ \cite{kumar2022magnetic, healey2022quantum}, to make heterostructures. The hBN spin defects were then used to characterize low dimensional magnetism in situ (Fig. \ref{Figure7:MagneticSensing}(a)). The Zeeman shift in the ODMR spectrum gives a direct measurement of the external field $B_{tot}$ which follows: $B_{tot}=\Delta f_{ESR}/\gamma_e$, where $\Delta f_{ESR}$ is the total frequency shift. Subtracting the bias field from the total field yields the field produced by the material of interest. hBN spin defects have a typical magnetic field detection sensitivity on the order of 10 $\mu$T/$\sqrt{\rm Hz}$ {\color{black} with a sensor volume of $\sim$0.007 $\mu$m$^3$}  (Table 1) \cite{gao2021high}, which is sufficient for studying 2D magnetic materials.
 
 Huang {\it et al.} performed a wide-field magnetic imaging using $V_B^-$ defects in an hBN/FGT vdW heterostructure with a wide-field microscope \cite{huang2022wide}.  The whole hBN flake contained $V_B^-$ spin defects  and the PL count rate from $V_B^-$ spin defects were spatially mapped onto a camera with single-photon sensitivity. By fitting the ODMR spectrum for each pixel of the camera, a map of the magnetic field was obtained (Fig. \ref{Figure7:MagneticSensing}(b)). Through a reverse-propagation protocols, Huang {\it et al.} reconstructed the corresponding magnetization $4\pi M$ (Fig. \ref{Figure7:MagneticSensing}(c)) \cite{huang2022wide}. 
 With this method, the magnetization near the whole flake was characterized  spatially as a function of temperature.
As shown in Fig. \ref{Figure7:MagneticSensing}(d), the measured FGT magnetization decreases with increasing temperature. Once the temperature went above the Curie temperature (225 K) of FGT, a significant suppression of the magnetization was observed. 

Healey {\it et al.} used similar approaches to image the magnetization of a 2D CrTe$_2$ flake in an  hBN/CrTe$_2$ heterostructure \cite{healey2022quantum}. They observed an in-plane magnetization of the 2D CrTe$_2$ flake, which gives a spontaneous magnetization $M_s$ $\approx$ 50 kA$\cdot$ m$^{-1}$. These experiments prove promising performance of hBN $V_B^-$ spin defects as nanoscale quantum sensors for studying low dimensional materials.

\begin{figure*}[htb]
	\centering
	{\resizebox*{14.3cm}{!}{\includegraphics{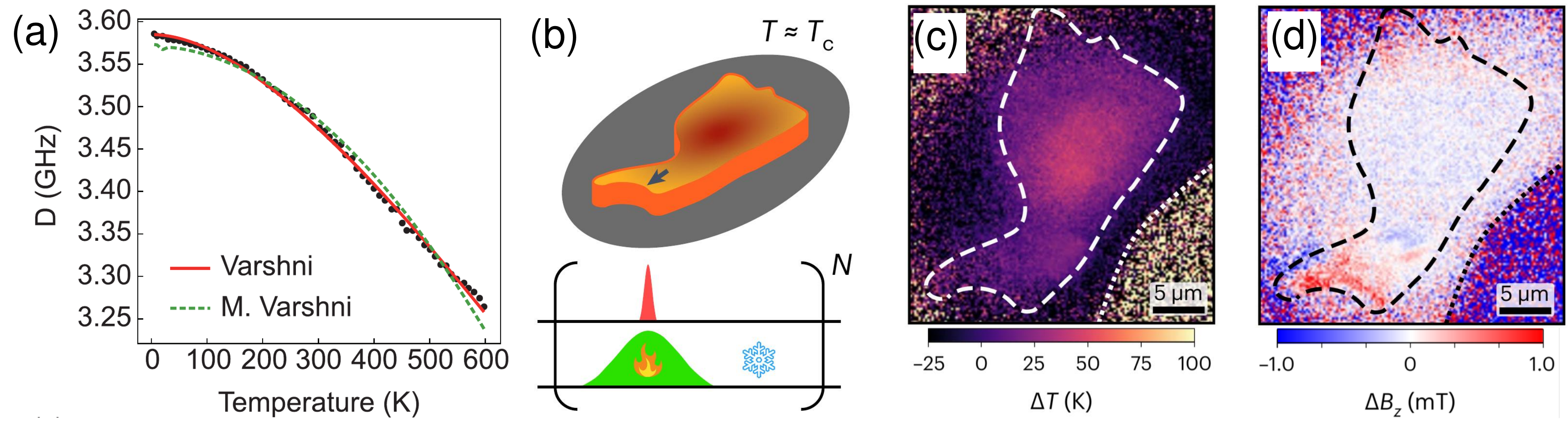}}}
	\caption{(a) Shift of the ground state ZFS parameter $D$ as a function of temperature. Reproduced from \cite{liu2021temperature}. Copyright by  American Chemical Society. (b) A pulse sequence showing heating a CrTe$_2$ flake with a laser pulse and probing its time-dependent temperature with hBN spin defects. (c) Temperature and (d) stray magnetic field map of the CrTe$_2$ flake near its Curie temperature. Reproduced from \cite{healey2022quantum}. Copyright by  Springer Nature. } \label{Figure8:temperature}
\end{figure*}

\begin{figure*}[htb]
	\centering
	{\resizebox*{14.3cm}{!}{\includegraphics{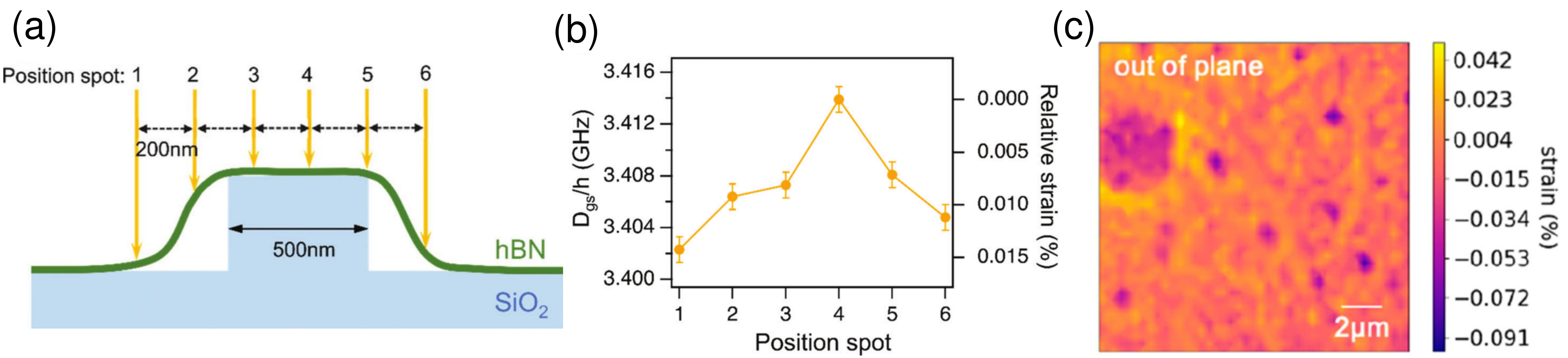}}}
	\caption{
		(a) Schematic showing an hBN nanosheet on a nanopillar. The distortion induces a position-dependent strain in the hBN nanosheet. (b) Measured value of $D_{gs}$ as a function of position on the pillar showing the strain. Reproduced from \cite{yang2022spin}. Copyright by Royal Society of Chemistry. (c) Measured strain map of a hBN flake with bubbles. Reproduced from \cite{lyu2022strain}. Copyright by American Chemical Society.
	} \label{Figure9:strain}
\end{figure*}

Besides measuring the DC magnetic field, hBN spin defects can also probe spin noises. Spin fluctuations in a magnetic material will generate high-frequency magnetic field noises. Such GHz magnetic noises will induce  transitions between different spin states,  and hence reduce the lifetime of nearby  hBN spin defects. By using the spin relaxometry measurement to measure the change of the spin lifetime of $V_B^-$ defects, the fluctuating magnetic fields of FGT near its Curie temperature was characterized  by Huang {\it et al.}  (Fig. \ref{Figure7:MagneticSensing} (e)) \cite{huang2022wide}.  This method will also be applicable to anti-ferromagnetic materials.

\subsection{Temperature sensing and imaging}

hBN has a high thermal conductivity (751 W/m$\cdot$K) \cite{cai2019high} and can be thinned down to a monolayer, which can significantly reduce its thermal capacity. These properties make hBN an ideal material platform for nanoscale temperature sensing. $V_b^-$ spin defects provide a way for temperature probing using hBN. The temperature affects $V_b^-$ defects via the longitudinal ZFS parameter $D_{gs}$. Gottscholl {\it et al.} \cite{gottscholl2021spin} and Liu \cite{liu2021temperature} {\it et al.} first reported the temperature dependence of the $V_B^-$ ODMR spectrum (Fig. \ref{Figure8:temperature}(a)). Over the temperature range from 295 K to 10 K, $D_{gs}$ undergoes a change of  $\Delta{D_{gs}}\approx 195$ MHz, which is approximately 30-fold larger than that of diamond NV centers. This temperature dependence is due to the change in the delocalization of the $V_B^-$ spin wave function under structure deformation. {\color{black} In an unoptimized experiment,} Gottscholl {\it et al.} estimated the temperature sensitivity of $V_B^-$ to be 3.82 K$/\sqrt{\rm Hz}$ \cite{gottscholl2021spin} {\color{black} with a sensor volume of $\sim$ 100 $\mu$m$^3$} (Table 1). 

In a temperature sensing experiment, Healey {\it et al.} performed time-resolved wide-field ODMR of  $V_B^-$ spin defects in a hBN/CrTe$_2$ heterostructure (Fig. \ref{Figure8:temperature})(b)-(d) \cite{healey2022quantum}. They mapped $\Delta$T under 
different laser illumination conditions and observed heating  due to laser absorption by the CrTe$_2$ flake (Fig. \ref{Figure8:temperature}(c)). By using a pulsed ODMR sensing protocol with a microwave pulse delayed by 200 ns, they found that the measured temperature decreased dramatically from 50 K to 15 K above the background.  Besides, Healey {\it et al.} also demonstrated thermal  imaging of an operating two-terminal graphene device with hBN spin defects. The device is made of a few-layer graphene ribbon covered with a 70-nm thick hBN flake. They recorded temperature maps with a constant current of $I$ = 3 mA flowing through the graphene. A temperature increase of $\Delta$T $\approx$ 6 K was observed as a consequence of Joule heating. These achievements prove the potential of hBN spin defects in nanoscale temperature sensing, which will be useful for the development of the next generation electronic devices.

\subsection{Strain sensing and imaging}
\begin{figure*}[hbt]
	\centering
	{\resizebox*{12cm}{!}{\includegraphics{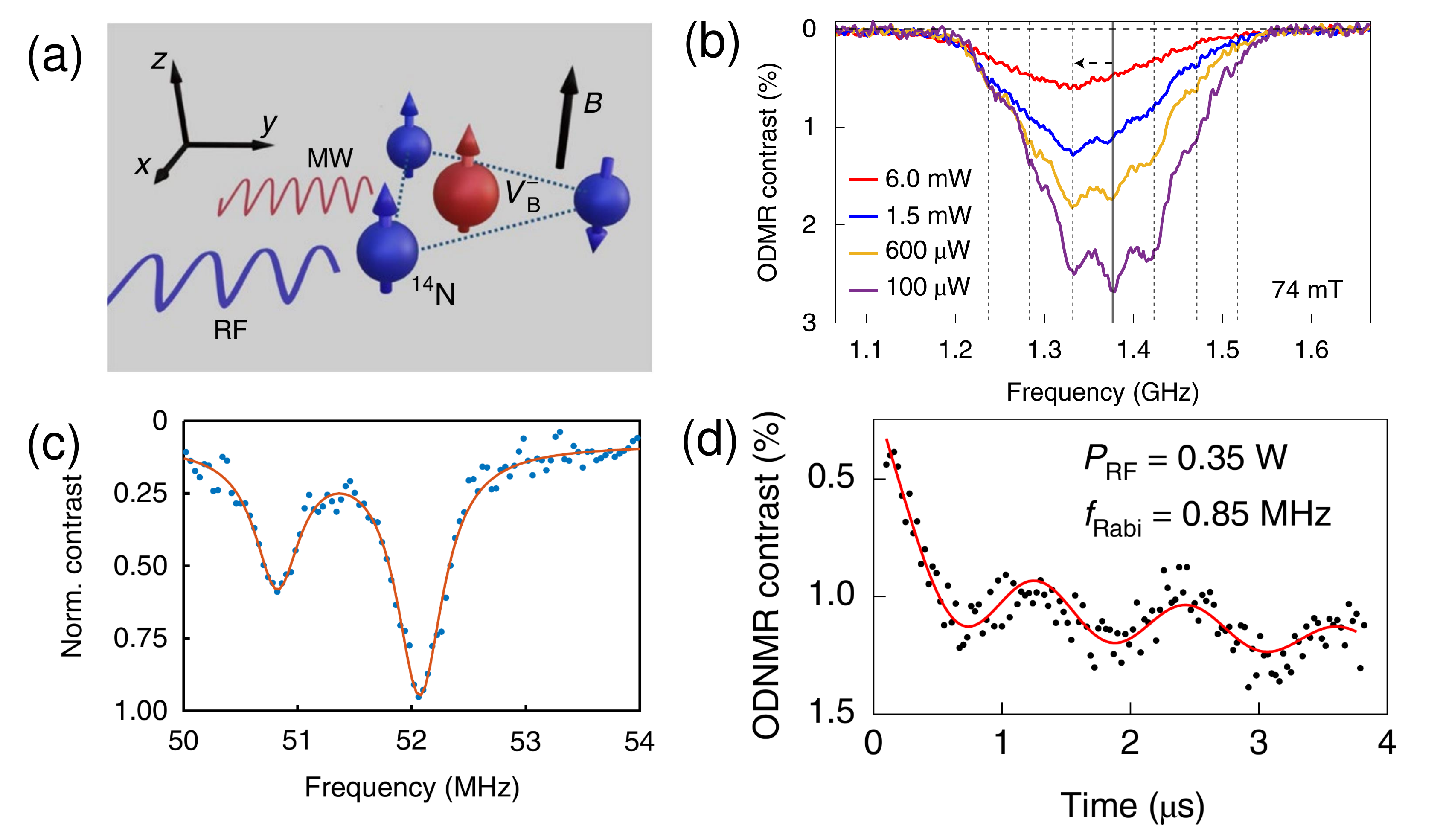}}}
	\caption{(a) Schematic of a $V_B^-$ electron spin surrounded by three nearest $^{14}N$ nuclei in magnetic field. A microwave field and an RF field are used to perform the ODNMR measurements. (b) Shift in the ODMR spectrum as a function of laser power indicating the nuclear spin  polarization (c) ODNMR contrast near the NMR transition frequency used to drive the nuclear spins. (d) Measured ODNMR contrast under a RF driving field showing a Rabi oscillation of nuclear spins. Reproduced from \cite{gao2022nuclear}. Copyright by Springer Nature.} \label{Figure10:NMR}
\end{figure*}

In situ measurement of the strain distribution of complex 2D heterostructures is highly desirable  for understanding many interesting phenomena in 2D systems. Fortunately, recent works on spin defects in hBN provide opportunities to satisfy these demands. In an experiment, the strain can be measured by $V_B^-$ defects via the change of ZFS parameters, which follow \cite{lyu2022strain}:
\begin{gather}
	h\nu_\pm = D_{gs} + D_S \pm \sqrt{E_{gs}^2 + E_S^2},\\
	D_S = a(\epsilon_{xx}+\epsilon_{yy})+b\epsilon_{zz},\\
	E_S=\sqrt{[\frac{c}{2}(\epsilon_{xx}-\epsilon_{yy})]^2+(c\epsilon_{xy})^2}.
\end{gather} 
where $\nu_\pm$ is the resonant frequency in the absence of an external magnetic field, and $\epsilon$ is the strain tensor. $D_{gs}$ and $E_{gs}$ are the original ZFS parameters without extra strain. $D_S$ and $E_S$ are additional ZFS parameters induced by the strain. $a$, $b$, and $c$ are the coupling parameters between the spin and the local strain. $a = -40.5$ GHz and $b = -24.5$ GHz are recently determined by Gottscholl {\it et al.} \cite{gottscholl2021spin}. Based on the change of ZFS parameters from ODMR spectra, one can extract the in-plane and out-of-plane strain fields.

Yang {\it et al.} measured the ODMR of spin defects in an hBN nanosheet transferred to a nanopillar array (Fig. \ref{Figure9:strain}(a)) \cite{yang2022spin}. By comparing the ODMR spectra  of spin defects on nanopillars and off nanopillars, they observed frequency shift due to different strains (Fig. \ref{Figure9:strain}(b)). In addition, they  found that introducing strain in the lattice helped increasing the emission intensity and ODMR contrast of the hBN spin defects. This may be explained by the symmetry breaking due to the strain field, which makes originally dipole forbidden transitions allowed.  Recently, correlative cathodoluminescence and  photoluminescence microscopes were used to reveal the effect of strain, which showed localized PL enhancement due to strain as well \cite{curie2022correlative}. Lyu {\it et al.} performed the wide-field ODMR to characterize the strain in the hBN lattice  (Fig. \ref{Figure9:strain}(c)) \cite{lyu2022strain}. The ZFS parameter $D_{gs}$ shifts with the strain on the lattice, and hence an in-plane strain ranging from -0.022$\%$ to 0.069$\%$, as well as an out-of-plane strain ranging from -0.091$\%$ to 0.042$\%$ were observed (Fig. \ref{Figure9:strain}(c)).

\subsection{Nuclear magnetic resonance}

Nanoscale nuclear magnetic resonance (NMR) is expected to have great impact on determining the structure of proteins or other complex materials. Also, nuclear spins often have longer coherence times than electron spins, which can be potentially used {\color{black} as} auxiliary memory qubits for improving the sensitivity in advanced pulsed sensing protocols. hBN is rich in nuclear spins, unlike a diamond that has only sparse nuclear spins. The $V_b^-$ defects in hBN are surrounded by three equivalent nitrogen atoms which consist of $I=1$ nuclear spins (Fig. \ref{Figure10:NMR}(a)). The hyperfine interaction between a $V_B^-$ electron spin and the nearest  three nitrogen nuclear spins results in seven hyperfine splittings in the ODMR spectrum as shown in Fig. \ref{Figure10:NMR}(b) \cite{gao2022nuclear,murzakhanov2022electron,liu2022coherent}. This hyperfine interaction enables the study of nuclear spins associated with the defect centers.

\begin{figure*}[htb]
	\centering
	{\resizebox*{10cm}{!}{\includegraphics{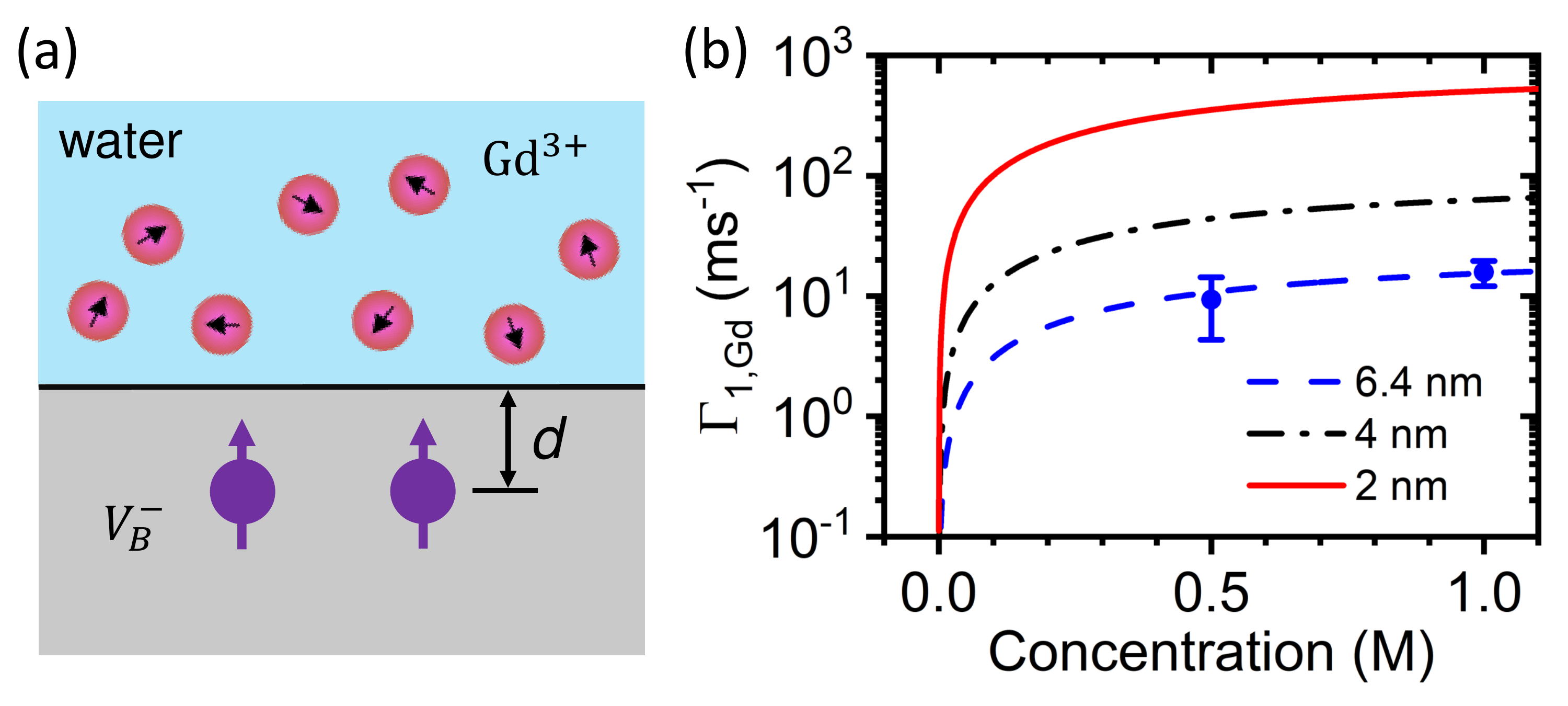}}}
	\caption{(a) Schematic of sensing paramagnetic ions in water with shallow $V_B^-$ spin defects with an average depth of $d$ in hBN. (b) Theoretical and experimental results of Gd$^{3+}$ induced spin relaxation rates as functions of Gd$^{3+}$ concentrations for $V_B^-$ spin defects at different depths. Reproduced from \cite{gao2023quantum}. } \label{Figure11liquid}
\end{figure*}

Optical polarization of the $^{14}$N nuclear spins up to $\approx$ 50$\%$ has been demonstrated in the $V_B^-$ system in hBN at room temperature\cite{gao2022nuclear}, mediated via the hyperfine interaction. After initialization of nuclear spins, a RF drive can be applied to observe NMR spectrum of these nuclear spins (Fig. \ref{Figure10:NMR}(c)). Multiple resonant dips were observed in the optically detected nuclear magnetic resonance (ODNMR) of the nuclear spins near the $V_B^-$ electron spin due to the electron spin-mediated strong coupling between nuclear spins. Coherent control and Rabi oscillations of the nuclear spins have been demonstrated (Fig. \ref{Figure10:NMR}(d)) with a nuclear spin dephasing time $T_{2n}^*$ of 3.5 $\mu s$ \cite{gao2022nuclear}. Theoretical investigations showed that it is possible to initialize the system into a high degree of hyperpolarization by exploiting the long coherence lifetimes of the nuclear spins using pairwise initialization of the electron-nuclear spin coupling \cite{tabesh2022active}. Isotopic enrichment of hBN with $^{10}$B has been shown to slightly improve the decoherence properties \cite{haykal2022decoherence}. These results show the potential of the $V_B^-$ spin defects for nanoscale NMR, and hBN nuclear spins for large scale quantum simulation \cite{cai2013large}.

{\color{black} 
\subsection{Detecting paramagnetic spins in liquids}

Paramagnetic ions and radicals, which possess at least one unpaired electron and consequently have non-zero electronic spins, play crucial roles in chemistry, biology, and medicine \cite{thomas2015breathing,griendling2016measurement}. They are involved in various physiological processes, including immune response, cell signaling, and metabolism.  To better understand their dynamics and functions in physiological processes, there is a growing need for nanoscale sensing and imaging of paramagnetic ions and radicals under ambient conditions. hBN with spin defects can be a promising sensor for this purpose owing to its small size, non-toxicity, and high stability in a broad range of environmental conditions.

Recently, Robertson {\it et al.} used $V_B^-$ spin defects in  hBN nanopowder to detect the Gd$^{3+}$ ions, a common paramagnetic contrast agent used in magnetic resonance imaging (MRI), under ambient conditions based on noise sensing protocols \cite{robertson2023detection}. They first put the hBN nanopowder on a microwave waveguide by drop casting. The bare spin defects in hBN nanopowder exhibited a $T_1$ relaxation time of 16.3$\pm$1.3 $\mu$s. After depositing a drop of gadolinium trichloride (GdCl$_3$) solution and letting it evaporate, the hBN nanoflakes were surrounded by GdCl$_3$ solids, leading to a significant reduction ($30\%$) of $T_1$ due to magnetic noise from Gd$^{3+}$ paramagnetic spins. Robertson {\it et al.} also tested the capability of hBN spin defects in a liquid environment. By comparing the $T_1$ of $V_B^-$ spin defects in hBN nanoflakes suspended in pure water and in a solution of 100 mM GdCl$_3$, a Gd$^{3+}$ ion induced relaxation rate of 10 $\pm$ 10 (ms)$^{-1}$ was observed \cite{robertson2023detection}. 
 
Gao et al. demonstrated the detection of magnetic noise from Gd$^{3+}$ ions in water using hBN $V_B^-$ spin defects within a microfluid structure (Fig. \ref{Figure11liquid}) \cite{gao2023quantum}. They transferred an hBN nanosheet doped with shallow spin defects (averaging 6.4 nm deep) onto a gold microwave waveguide transmission line and sealed it in a microfluid structure for liquid changes during measurements.  By using deionized water as the reference,  the additional relaxation rates induced by Gd$^{3+}$ ions are found to be $\Gamma_{1,Gd}$ = 15.8 $\pm$3.8 (ms)$^{-1}$ for 1 M Gd$^{3+}$ ions and 9.4 $\pm$ 5.0 (ms)$^{-1}$ for 0.5 M Gd$^{3+}$ ions, which agreed well with theoretical predictions (Fig. \ref{Figure11liquid}) \cite{gao2023quantum}. In addition, by using CW ODMR technique, Gao {\it et al.} also showed the  contrast reduction of the ODMR spectra when the spin defects were in the presence of Gd$^{3+}$ ions in water \cite{gao2023quantum}.  Such a reduction depends on the paramagnetic ion concentration, which provides an alternative, simpler method to detect paramagnetic ions in liquids. 

}

\begin{figure*}[htb]
	\centering
	{\resizebox*{14.3cm}{!}{\includegraphics{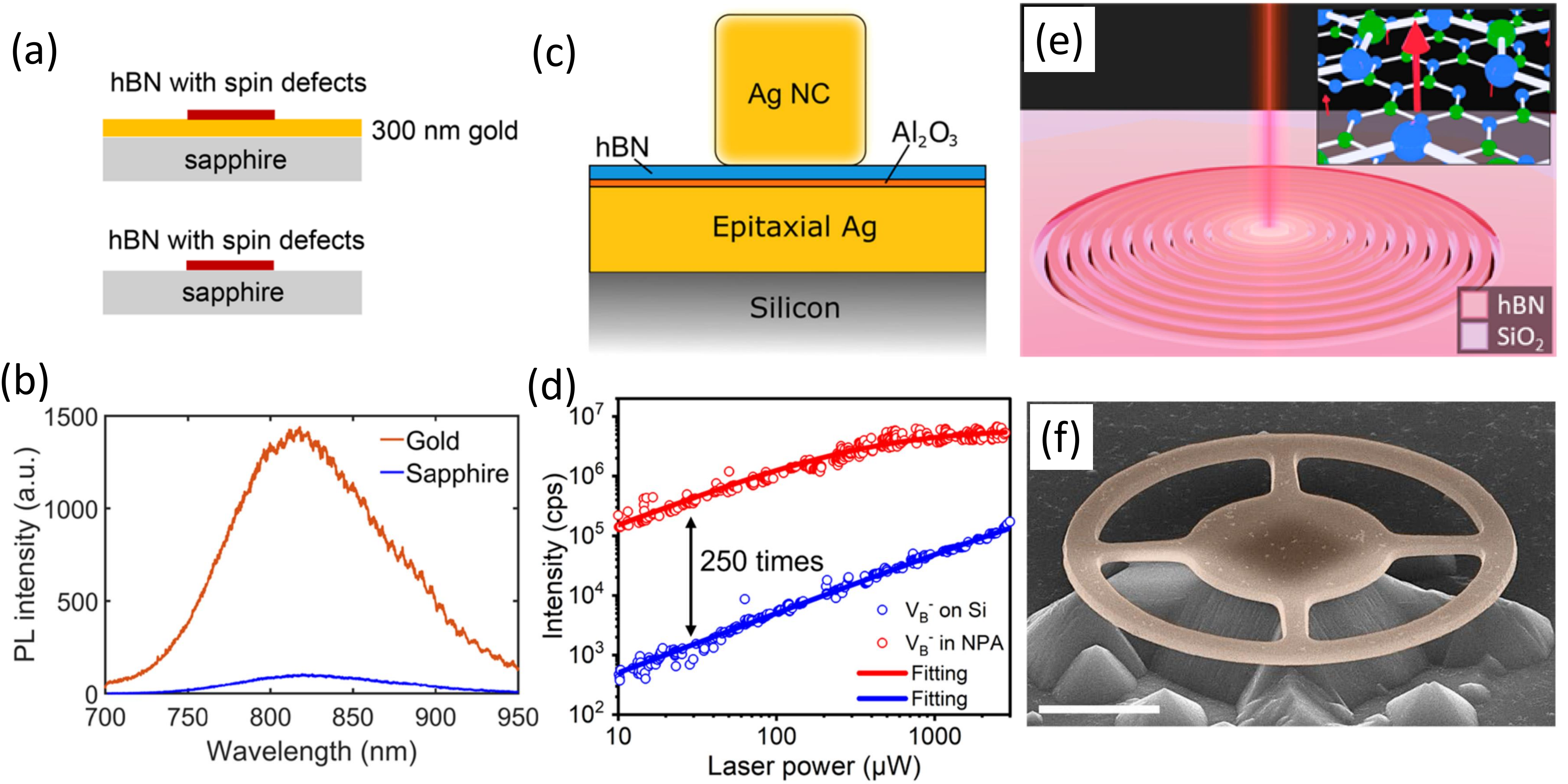}}}
	\caption{(a) Schematic of an hBN fake deposited on a gold-coated sapphire substrate or on a bare sapphire substrate. (b) Increased photoluminescence intensity from hBN $V_B^-$ defects deposited on a gold-coated sapphire substrate. Reproduced from \cite{gao2021high}. Copyright by American Chemical Society. (c) Schematic of a plasmonic nanocavity formed between a silver nanocube and an expitaxial silver film separated by ultrathin hBN and Al$_2$O$_3$ thin films. (d) Increased photoluminescence intensity as a function of laser power for hBN spin defects in a silver nanocavity. Reproduced from \cite{xu2022greatly}. Copyright by American Chemical Society. (e) Schematic of a bullseye cavity fabriacted out of hBN with the emission being couple out of plane for maximum collection. Reproduced from \cite{froch2021coupling}.  Copyright by American Chemical Society. (f) SEM image of a hBN/TiO$_2$ ring resonator fabricated to increase emission. Reproduced from \cite{nonahal2022coupling}. Copyright by Royal Society of Chemistry. Scale bar: 2 $\mu$m. } \label{Figure12:brightness}
\end{figure*}

\subsection{RF signal sensing}

Spin defects in hBN  provide a promising tool for ac magnetic field sensing. By utilizing a dynamic decoupling (DD) sequence, one can not only extend the spin coherence time \cite{gong2022coherent,ramsay2023coherence}, but also detect an ac magnetic field, such as a radiofrequency (RF) signal \cite{rizzato2022extending}. Rizzato {\it et al.}  used both pulsed DD (XY8-N) and continuous DD (spinlock) techniques to {\color{black} sense} RF signals \cite{rizzato2022extending}. The pulsed DD sequence acts like a narrow-band RF filter and the $V_B^-$ superposition state accumulates a maximal  phase if the free evolution time $\tau$ matches $\tau=1/(4\nu_{RF})$, where $\nu_{RF}$ is the frequency of the RF signal \cite{sasaki2016broadband}. This will further lead to a dip in the PL intensity during the optical readout. However, the detectable RF frequency in this case is limited by the decoherence rate, and this sensing protocol becomes less effective when $\nu_{RF} <10$ MHz. Rizzato further demonstrated RF sensing with arbitrary frequency resolution by implementing the coherently averaged synchronized readout (CASR) scheme \cite{glenn2018high,boss2017quantum,schmitt2017submillihertz}. The basic idea is to use a train of DD sequence synchronized with the RF signal. If $\nu_{RF}\ne 1/4\tau$, then a consecutive phase will be accumulated. As a result, the PL will oscillate at a rate equal to the difference between $\nu_{RF}$ and $1/4\tau$. With such a technique, sensing RF frequencies with sub-Hz frequency resolution is doable, despite the intrinsically short coherence time of $V_B^-$ defects.

\section{Improving sensitivity}
Despite the great potential of hBN spin defects, the sensitivity of $V_B^-$ spin defects is currently limited by their low brightness and short coherence time. {\color{black} A comparison of sensitivities between hBN $V_B^-$ defects and other spin defects in bulk material systems can be found in  references \cite{liu2021temperature,gottscholl2021spin}.}   Researchers have developed various methods to improve the PL count rates and extend spin coherence. In this section, we review the techniques that have been used for improving the sensitivity of $V_B^-$ spin defects.

\subsection{Improving brightness}

The low brightness of hBN $V_B^-$ defects has resulted in the fact that single $V_B^-$ defects have not yet been observed in this system. {\color{black} Small ensembles of $V_B^-$ contained in a 0.01 $\mu$m$^3$ volume are measurable but are relatively dim, comparable to the signal from a single diamond NV center.}  Therefore, improving  the photon emission rate and the photon collection efficiency is a major task for practical quantum sensing applications. A number of approaches have been explored to improve the brightness of $V_B^-$ defects. A major idea is to couple the spin defects to plasmonic structures, where the intense localized electromagnetic field  increases the excitation and spontaneous emission rate of  spin defects \cite{koenderink2017single,pelton2015modified,tame2013quantum}. Nanophotonic structures can also be used to change the emission pattern of spin defects to improve photon collection. Increasing the size of a spin ensemble can also increase the brightness, at the expense of reducing spatial resolution. 

\begin{figure*}[htb]
	\centering
	{\resizebox*{12cm}{!}{\includegraphics{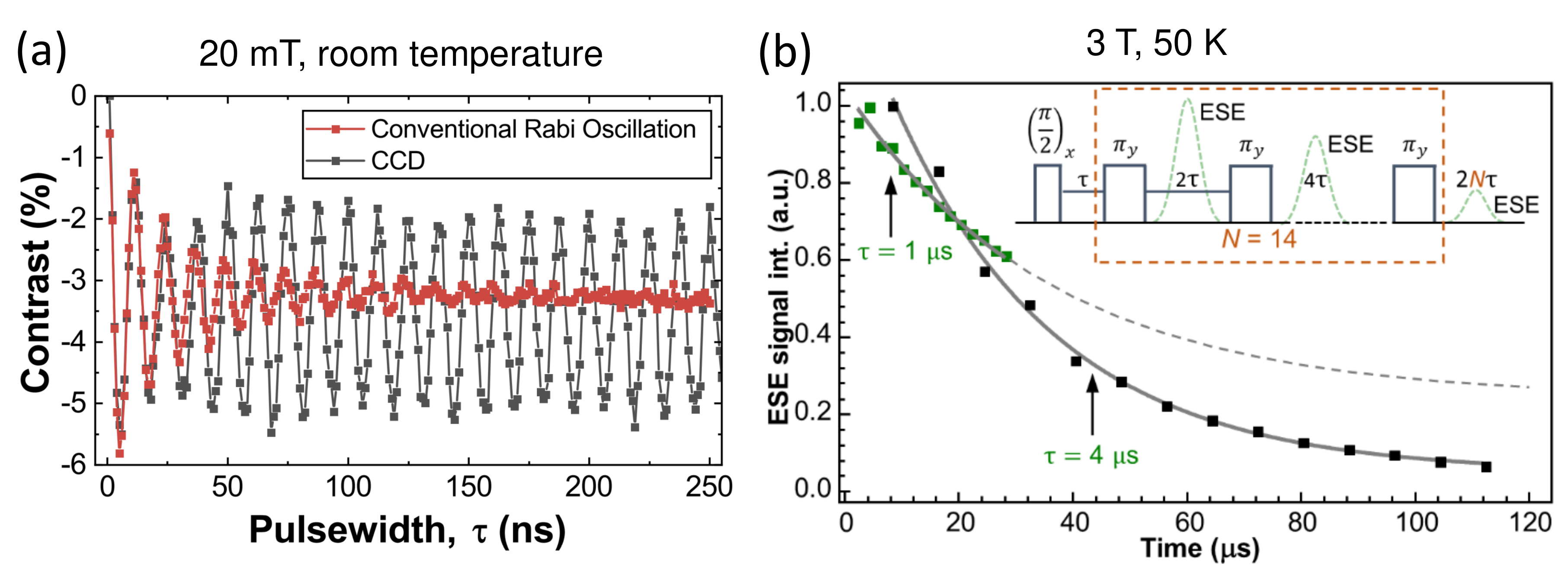}}}
	\caption{(a) Stabilization of Rabi oscillation by using an amplitude modulated CCD scheme in a weak (20 mT) magnetic field at room temperature. The spin dephasing time is  increased from 31 ns to 2.2 $\mu$s. Reproduced from \cite{ramsay2023coherence}.  (b)  Reducing the decay of the electron spin echo (ESE) signal in a 3 T magnetic field at 50 K using the CPMG scheme with two different pulses lengths. The coherence time was extended to 36 $\mu$s for the gray dashed line ($\tau=1 \mu s$) and to 28.6 $\mu$s for the solid line ($\tau=4 \mu s$). Reproduced from \cite{murzakhanov2022electron}. Copyright by American Chemical Society. } \label{Figure13}
\end{figure*}

Gao {\it et al.} used localized surface plasmons on a gold-film microwave waveguide to increase the brightness of $V_B^-$ defects, and observed an up to 17-fold PL enhancement  (Fig. \ref{Figure12:brightness}(a),(b)) \cite{gao2021high}. The gold film showed a roughness at the nanoscale, which
supports localized surface plasmons that increases photon emission rate by the Purcell effect. Meanwhile, the gold microwave waveguide generates a strong in-plane microwave magnetic field that can efficiently drive electron spin transitions. As a result,  an up to $46\%$ ODMR contrast was observed at room temperature \cite{gao2021high}. Such a high ODMR contrast also improves the sensitivity of spin defects.      Gapped plamonic nanocavities can further improve the brightness \cite{xu2022greatly, mendelson2022coupling}.  Xu {\it et al.}  drop-casted silver nanocubes onto hBN flakes which were transferred on an alumina-coated epitaxial silver film (Fig. \ref{Figure12:brightness}(c)). The silver cubes and the metallic surface formed nanopatch antennas, where the PL intensities were enhanced by up to 250 times  while maintaining the ODMR contrast (Fig. \ref{Figure12:brightness}(d)). As the silver nanocube was much smaller than the optical diffraction limit, some collected photons came from spin defects not under the silver cube. The corrected PL enhancement is more than 1685 times for spin defects under the silver nanocube \cite{xu2022greatly}.

Besides plasmonic structures, nanophotonic structures have also been developed to improve the brightness of hBN spin defects. Froch  {\it et al.} integrated the hBN $V_B^-$ defects into a bullseye cavity (Fig. \ref{Figure12:brightness}(e)) and achieved an enhancement factor of 6.5 \cite{froch2021coupling}. Nonahal {\it et al.} fabricated high quality TiO$_2$ ring resonators (Fig. \ref{Figure12:brightness}(f)) \cite{nonahal2022coupling}. By coupling $V_B^-$ defects into the whispering gallery modes of the ring resonators, Nonahal {\it et al.} increased the collected PL by a factor of 7. Reflective dielectric cavity structures have been reported by Zeng {\it et al.} \cite{zeng2022reflective}. Zeng {\it et al.} used a metal reflective layer under the hBN flakes, filled with a SiO$_2$ dielectric layer in the
middle. By optimizing the thickness of the dielectric layer, a PL enhancement factor of 7 was achieved. 

Twisting hBN may be a potential way to improve the brightness of $V_B^-$ defects. This technique has been implemented by Su {\it et al.} for another type of hBN color centers and was shown to be able to increase the emission by two orders of magnitudes \cite{su2022tuning}.

\subsection{Extending coherence time}

Another promising approach to increase the sensitivity of hBN spin defects is to extend their spin coherence time $T_2$, which will allow more powerful pulsed sensing protocols. 
The measured coherence times of $V_B^-$ defects in  hBN \cite{gottscholl2021room,ye2019spin} are shorter than those of diamond NV centers \cite{stanwix2010coherence} due to the high concentration of nuclear spins in hBN. The $T_2^*$ of $V_B^-$ defects was reported to be around 100 ns, which limits their application in quantum sensing, computing and simulation. There has been some progress in increasing the coherence times via pulse sequences. Ramsay {\it et al.} extended the coherence time of $V_B^-$ Rabi oscillations to as high as 4 $\mu$s at room temperature by using  amplitude modulated concatenated dynamic decoupling (CCD)  \cite{ramsay2023coherence}. This techniques utilized a strong continuous microwave drive along with amplitude modulation to stabilize the Rabi oscillation dramatically (Fig. \ref{Figure13}(a)). The protected superposition qubit shows an up to 800 ns coherence time, which is much better than that of the unprotected $V_B^-$ spins. Similarly, Gong {\it et al.} reported coherent times of up to 500 ns using a DROID (Disorder-RObust Interaction-Decoupling)  protocol which is particularly effective for strongly interacting electron spin systems \cite{gong2022coherent}. There are theoretical proposals to increase the $T_2$ by isotopic purification using $^{10}$B atoms during material growth \cite{lee2022first}. This has been demonstrated experimentally \cite{haykal2022decoherence}, although the observed improvement was minor (from 46 ns to 62 ns) so far.  Reducing the temperature can increase the spin lifetime $T_1$, which can also benefit $T_2$. In a strong magnetic field ($\sim$3.2 T) and a cryogenic temperature (50 K), the $T_2$ of $V_B^-$ electron spins has been extended to  36 $\mu$s using the Carr-Purcell-Meiboom-Gill (CPMG) pulse sequence in a W-band ($\sim 94$ GHz) commercial pulsed ESR spectrometer (Fig. \ref{Figure13}(b)) \cite{murzakhanov2022electron}.

\section {Conclusion and outlook}

In this review, we discussed quantum sensing and imaging using hBN spin defects (e.g., $V_B^-$ defects), which have promising future for various applications. We introduced the basic properties of spin defects in hBN and  techniques to create them. We also discussed different sensing protocols that had been applied to  hBN spin defects. The recent progress in utilizing hBN $V_B^-$ spin defects for detecting 2D magnetism, temperature, strain,  nuclear spins, and RF signals is summarized. Finally, we review  methods to enhance the PL intensity and extend the coherence time to improve the sensitivity of hBN spin defects. 

Some other applications of hBN spin defects are emerging. Recently, an array of V$_B^-$ spin defect spots with a size of $(100 {\rm nm})^2$ were generated  using a helium ion microscope with nanoscale precision \cite{sasaki2023magnetic}. By assembling data from multiple spin defect spots together, the magnetic field induced by the current in a wire was visualized with a spatial resolution beyond the diffraction limit in one direction \cite{sasaki2023magnetic}. In the future, stimulated emission depletion (STED) microscopy \cite{khatri2021stimulated} 
may enable hBN spin defects to image samples beyond the diffraction limit in two directions.  
For wide-field quantum sensing and imaging, it will  be important to generate a large homogeneous microwave field to drive spin defects in a large area. Recently, a microwave double arc resonator for efficient transferring of the microwave field at 3.8 GHz was designed and fabricated to drive hBN spin defects efficiently over a large field of view \cite{tran2023coupling}. {\color{black} It has been proposed that hBN spin defects could have applications in high-harmonic generation \cite{mrudul2020high}.
Thanks to the small mass of an hBN nanosheet, an electron spin can have a large effect on the mechanical motion of an hBN resonator. Thus hBN spin defects will have important applications in spin optomechanics \cite{abdi2017spin,abdi2019quantum,shandilya2019hexagonal,hu2022stability,yazdi2023spin}. }

{\color{black} For all these future applications, efforts to optimize performance of hBN spin defect sensors are highly warranted. In addition to improving brightness and coherence based on reported methods, replacing $^{14}$N by $^{15}$N isotope may help narrow down the ODMR spectral linewidth and further enhance sensitivity.  Further investigations are necessary to gain a deeper understanding of the physical properties of $V_B^-$ spin defects, including quantum efficiency, charge dynamics, and coherent dynamics under various sample generation conditions. } 

Thanks to their 2D geometry, hBN spin defects will be well-suited for nanoscale  NMR and other applications in surface chemistry and biology. Besides $V_B^-$ spin ensembles, other single spin defects in hBN may provide higher sensitivity and spatial resolution. 
The sensing protocols developed for hBN spin defects will also be applicable for optically addressable spin defects to be discovered in other 2D materials (e.g. WS$_2$, MoSe$_2$) \cite{li2022carbon,lee2022spin}.  
Beyond quantum sensing, spin defects in 2D materials will  have applications in quantum simulation \cite{cai2013large,hu2022stability} and quantum communication \cite{atature2018material,azzam2021prospects,kubanek2022review}. 

\vspace*{0.1cm}

\section* {Acknowledgments}
T.L. acknowledges supports from the the DARPA ARRIVE program, the National Science Foundation under grant no. PHY-2110591, and the Purdue Quantum Science and Engineering Institute through a seed grant. I.A. acknowledges the Australian Research Council (CE200100010, FT220100053) and the Office of Naval Research Global (N62909-22-1-2028) for the financial support.



%

\end{document}